\documentclass[
    reprint,
    superscriptaddress,
    floatfix,
    preprintnumbers,
    longbibliography,
    prd,
    amsmath,
    amssymb,
    aps,
    nofootinbib
]{revtex4-2}

\usepackage{graphicx}
\usepackage{dcolumn}
\usepackage{bm}

\RequirePackage{graphicx}
\usepackage{xcolor}
\usepackage{booktabs}
\usepackage{amsmath}
\usepackage{amssymb}
\usepackage{relsize}
\usepackage{makecell}
\usepackage{multirow}
\usepackage[version=4]{mhchem}
\usepackage{textgreek}
\usepackage{url}
\usepackage{afterpage}

\usepackage{hyperref}
\usepackage{soul}

\usepackage{lineno}

\usepackage{comment}

\usepackage{floatrow}
\newfloatcommand{capbtabbox}{table}[][\FBwidth]

\usepackage{makecell}
\usepackage{cellspace} %
\usepackage{letltxmacro}
\LetLtxMacro{\oldcite}{\cite}
\renewcommand{\cite}[1]{\mbox{\oldcite{#1}}}
\newcommand{\Qrec}{Q_{\rm rec}}
\newcommand{\Qdep}{Q_{\rm dep}}

\begin{document}

\preprint{APS/123-QED}


\title{Opportunities and challenges to study solar neutrinos with a Q-Pix pixel readout}

\newcommand{\Harvard}{Department of Physics, Harvard University, Cambridge, MA 02138, USA}
\newcommand{\Arlington}{Department of Physics, University of Texas at Arlington, Arlington, TX 76019, USA}
\newcommand{\Manchester}{Department of Physics and Astronomy, University of Manchester, Manchester M13 9PL, UK} 
\newcommand{\Imperial}{Department of Physics, Imperial College London, London SW7 2AZ, UK} 
\newcommand{\Wellesley}{Department of Physics and Astronomy, Wellesley College, Wellesley, MA 02481, USA}
\newcommand{\Fermilab}{Fermi National Accelerator Laboratory, Batavia, IL 60510, USA}
\newcommand{\LBNL}{Lawrence Berkeley National Laboratory, Berkeley, CA 94720, USA}
\newcommand{\ORNL}{Oak Ridge National Laboratory, Oak Ridge, TN 37831, USA}
\newcommand{\ANL}{Argonne National Laboratory, Lemont, IL 60439, USA}
\newcommand{\Hawaii}{Department of Physics and Astronomy, University of Hawaii, Honolulu, HI 96822, USA}
\newcommand{\UPenn}{Department of Physics and Astronomy, University of Pennsylvania, Philadelphia, PA 19104, USA}
\newcommand{\Edinburgh}{School of Physics and Astronomy, University of Edinburgh, Edinburgh EH8 9YL, UK}
\newcommand{\PNNL}{Pacific Northwest National Laboratory, Richland, WA 99354}
\newcommand{\Sheffield}{School of Mathematical and Physical Sciences, University of Sheffield, Sheffield, S3 7RH, UK}

\author{M.\'A~Garc\'ia-Peris*}
\affiliation{\Manchester}
\thanks{Corresponding author: \url{miguel.garciaperis@manchester.ac.uk}}
\author{G.~Ruiz}
\affiliation{\Manchester}
\author{S.~Kubota}
\affiliation{\Manchester}
\affiliation{\Harvard}
\affiliation{\LBNL}
\author{A.~Navrer-Agasson}
\affiliation{\Imperial}
\author{G.~V.~Stenico}
\affiliation{\Edinburgh}
\author{E.~Gramellini} 
\affiliation{\Manchester}
\author{R.~Guenette}
\affiliation{\Manchester}
\author{J.~Asaadi}
\affiliation{\Arlington}
\author{J.B.R.~Battat} 
\affiliation{\Wellesley}
\author{V. A.~Chirayath}
\affiliation{\Arlington}
\author{E.~Church}
\affiliation{\PNNL}
\author{Z.~Djurcic}
\affiliation{\ANL}
\author{A. C.~Ezeribe}
\affiliation{\Sheffield}
\author{J. N.~Gainer}
\affiliation{\Arlington}
\author{G.~Gansle}
\affiliation{\Arlington}
\author{K.~Keefe}
\affiliation{\Arlington}
\author{N.~Lane}
\affiliation{\Manchester}
\author{C.~Mauger}
\affiliation{\UPenn}
\author{Y.~Mei}  
\affiliation{\Arlington}
\author{F.M.~Newcomer}
\affiliation{\UPenn}
\author{D.R.~Nygren}
\affiliation{\Arlington}
\author{M.~Rooks}
\affiliation{\Arlington}
\author{P.~Sau}
\affiliation{\Arlington}
\author{O.~Seidel}
\affiliation{\Arlington}
\author{S.~Söldner-Rembold}
\affiliation{\Imperial}
\author{I.~Tzoka}
\affiliation{\Arlington}
\author{R.~Van~Berg} 
\affiliation{\UPenn}
\date{\today}
\begin{abstract}

The study of solar neutrinos presents significant opportunities in astrophysics, nuclear physics, and particle physics. However, the low-energy nature of these neutrinos introduces considerable challenges to isolate them from background events, requiring detectors with low-energy threshold, high spatial and energy resolutions, and low data rate. 
We present the study of solar neutrinos with a kiloton-scale liquid argon detector located underground, instrumented with a pixel readout using the Q-Pix technology.  We explore the potential of using volume fiducialization, directional topological information, light signal coincidence and pulse-shape discrimination to enhance solar neutrino sensitivity.  We find that discriminating neutrino signals below 5 MeV is very difficult. However, we show that these methods are useful for the detection of solar neutrinos when external backgrounds are sufficiently understood and when the detector is built using low-background techniques.  When building a workable background model for this study, we identify  $\gamma$ background from the cavern walls and from capture of $\alpha$ particles in radon decay chains as both critical to solar neutrino sensitivity and significantly underconstrained by existing measurements.
Finally, we highlight  that the main advantage of the use of Q-Pix for solar neutrino studies lies in its ability to enable the continuous readout of all low-energy events with minimal data rates and manageable storage for further offline analyses.

\end{abstract}

\maketitle
\section{Introduction}\label{sec:intro}

The Standard Solar Model (SSM)~\cite{SolarModel_orig} describes the mechanisms of energy production in the Sun and, by extension, of other stars, as well as the fundamentals of their life cycle and the generation of elements in the universe. The SSM predicts the internal solar structure and the solar neutrino flux~\cite{Bahcall:2000nu}.

Measurements of the Sun's metallicity, which refers to the fraction of the solar mass consisting of elements heavier than helium, can constrain the Sun's evolutionary history. The metallicity predicted by the SSM was found to be in good agreement with past observations, such as those from the GONG (Global Oscillations Network Group) project~\cite{GONG}. However, more recent measurements show a different result~\cite{metallicity_disagreement_1, metallicity_disagreement_2}. This discrepancy is now known as the solar metallicity problem.

Solar neutrinos, produced in the center of the Sun during nuclear fusion reactions, can be a rich testing ground of the SSM. Especially, the precise measurement of the flux of neutrinos from the CNO (carbon-nitrogen-oxygen) cycle can provide a solution to the solar metallicity problem, as higher metallicity leads to higher CNO neutrino production. Borexino published the first measurements on the CNO flux~\cite{Borexino_2023}, but more measurements are needed. Furthermore, neutrinos produced through the rare high-energy process ($\ce{^{3}He} + p \rightarrow \ce{^{4}He} + e^{+} + \nu_{e}$), also known as hep neutrinos, have been predicted by the SSM but have not yet been observed. As hep neutrinos are the most energetic solar neutrinos, they can also offer a unique probe to the higher energy regime of solar models. Overall, solar neutrino research has broad implications for astrophysics and cosmology, as a better understanding of the nuclear reactions occurring in the Sun can extend our understanding and refine our models of stellar evolution and nucleosynthesis~\cite{Fischer:2023ebq}. 

Solar neutrinos played a crucial role in the discovery of neutrino oscillations~\cite{Bahcall:2000nu, SNO_oscillation}. Detailed studies of solar neutrinos can also offer precision measurements in neutrino oscillation physics. Solar neutrinos with the second highest energy spectrum are produced the most abundantly through the $\beta$ decay of boron-8 (B8) in the Sun's core. Their detection was historically important as a resolution to the solar neutrino problem~\cite{SNO_solar_problem, Sage_solar_problem}.

Mikheyev–Smirnov–Wolfenstein (MSW) matter effects have significant impact on the propagation of neutrinos through the dense medium of the Sun~\cite{Wolfenstein:1977ue}, making B8 neutrinos a powerful tool to measure neutrino oscillation parameters, including mass-squared differences and mixing angles~\cite{Maltoni:2015kca}. The long-standing tension between the $\Delta m^2_{21}$ measured by KamLAND and solar neutrino experiments is of particular interest, as the Kamland finds a best fit 1.5 standard deviations larger than the solar neutrino experiments~\cite{bib:kamland,bib:vs_kamland}. Hence, the development of a next-generation neutrino experiment capable of detecting solar neutrinos is of significant interest across various disciplines. 

Studying solar neutrinos presents specific experimental challenges. Despite the abundant flux of solar neutrinos~\cite{Bahcall:2004qv}, the small neutrino interaction rates at lower energies and the resulting low-energy signal events in detectors make their detection difficult. The energy spectrum of solar neutrinos peaks around $10$~MeV for the B8 and hep processes, which is significantly lower than the energy range of accelerator neutrinos (ranging from tens of MeVs to GeVs) and even lower than astrophysical supernova neutrinos, which have energies up to approximately $50$~MeV~\cite{SN_energy_spectrum}. The particles produced in solar neutrino interactions result in low-energy events that require detectors with sufficiently low energy-detection thresholds, excellent tracking capabilities, and high reconstruction efficiencies. Radioactive and external background rates in the $1$--$10$~MeV range can be high in experiments not specifically designed for rare event searches. Therefore, background mitigation strategies will be crucial to ensure the successful study of solar neutrinos.

Liquid-argon Time Projection Chambers (LArTPCs) have been widely used in neutrino experiments like ICARUS~\cite{ICARUS}, MicroBooNE~\cite{MicroBooNE:2016pwy}, and several other experiments~\cite{LArIAT, ArgoNeuT, SBND}. The common technology is a wire-based readout, where ionization charge is detected using consecutive layers of wire planes, resulting in a projective readout. In projective readouts two of the three spatial coordinates are obtained by interpolation of the charge signal on the wire planes, while the absolute component along the drift direction is calculated by matching the charge deposited with the light signals. However, projective readouts is subject to reconstruction ambiguities for events with complex, dense topologies~\cite{Adams:2019uqx}. By capturing spatial dimensions as inherently 3D information, a pixelated readout eliminates such ambiguities associated with 2D projections. Additionally, the long sense wires in a projective readout system add capacitive load to the readout electronics, creating noise that can limit the energy detection threshold~\cite{MicroBooNE:2017qiu}. The underlying projective concept of LArTPCs requires that the data is readout continuously during the entire time that the ionization charge from an event drifts towards the wire planes, resulting in large data rates. Finally, in the next generation of experiments, where the wires are mounted on large support structures~\cite{bib:DUNETDR4}, the strict engineering requirements make these plane assemblies complex and costly to manufacture. 

A way to reduce the technical complexity of wire readouts is to use Charge Readout Planes (CRPs)~\cite{bib:vd_tdr}. These are mechanically robust, modular for easy assembly, and suitable for mass production. In this approach, the anodes are composed of stacked, segmented, and perforated printed circuit boards (PCBs) with etched electrodes. Even in the case of CRPs, the readout is still intrinsically projective, and the limitations due to reconstruction ambiguities or the problem with the high capacitance are not mitigated. A 3D readout made of pixels could address the limitation of projective readouts maintaining mechanical robustness. Pixel readouts deployed in a 10-kton LArTPCs require $\mathcal{O}(10^{8})$ readout channels, posing challenges in terms of the data rate and the power consumption. Research collaborations worldwide, such as LArPix~\cite{Dwyer:2018phu, LArPix_2x2},  SoLAr~\cite{SoLAr:2024fwt} and Q-Pix~\cite{Q-Pix_org}, have been working to enable large-scale LArTPCs equipped with fully pixelated low-power charge readout.

In this paper, we study the capabilities of a generic large LArTPC, instrumented with the Q-Pix readout technology~\cite{Q-Pix_org}, located underground to detect solar neutrinos. We assess the impact of an extensive set of internal and external backgrounds for two detector scenarios: a multi-kt LArTPC module not optimized for low-background (\textit{high background scenario}) and a module specifically designed to reduce backgrounds as outlined in the SLoMo detector proposal~\cite{bib:slomo} (\textit{low-background scenario}).

\section{Experimental design assumptions} 
\noindent We outline here our assumptions regarding the experimental setup under study, the detector characteristics, and the host cavern. 

\subsection{Detector}

\begin{figure}[htbp]
\centering
\includegraphics[width=0.9\textwidth]{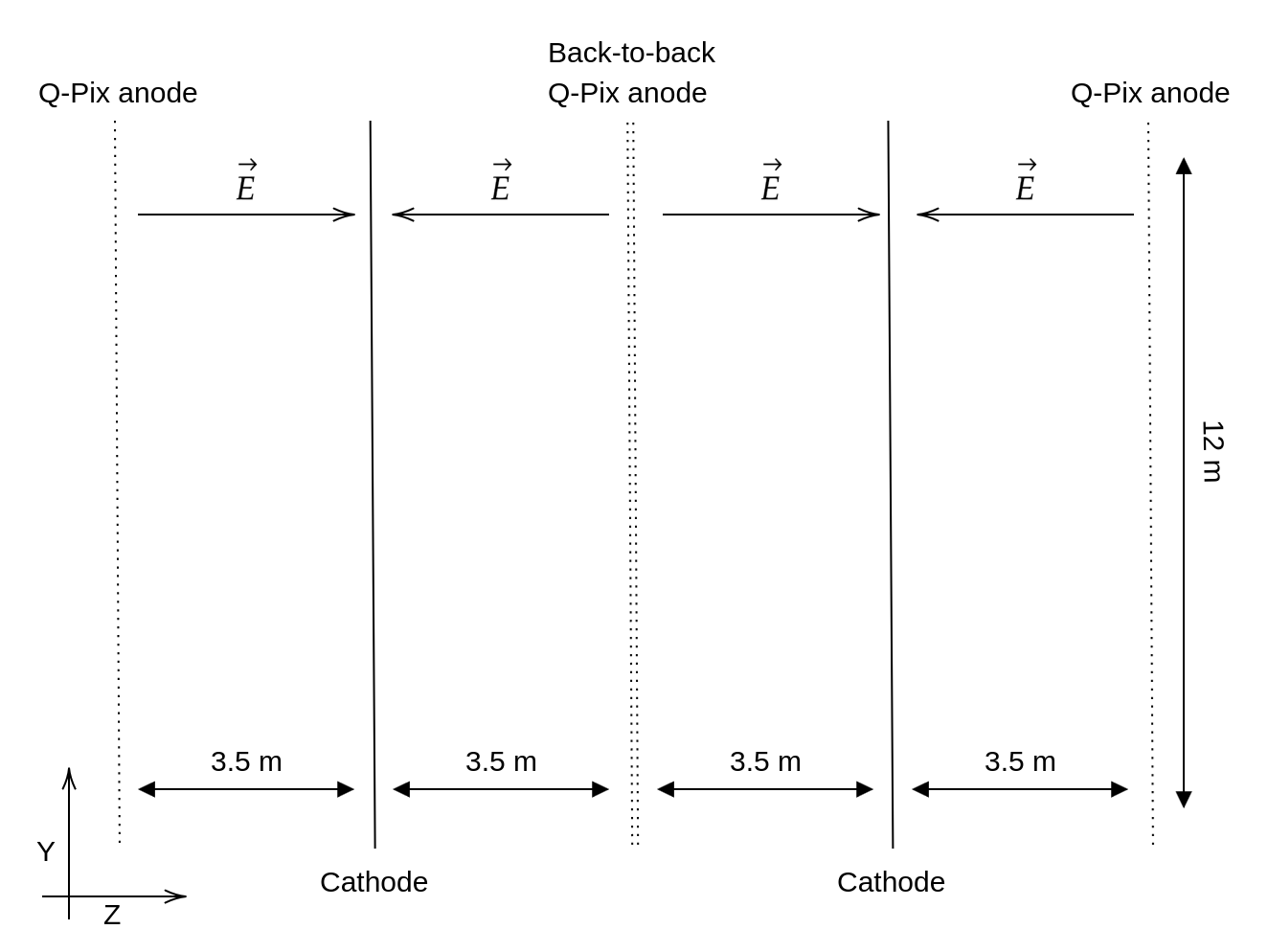}
\caption{Schematic cross-section view of the detector model with four drift volumes of 3.5 m each. The innermost Q-Pix anode plane consists of two pixel planes. The z axis runs parallel to the planes, one facing each drift volume. Not to scale. \label{fig:detector}}
\end{figure}

\noindent \textit{High Background Scenario.} In the high background scenario, we consider a 17.5~kiloton LArTPC with 10~kiloton fiducial volume filled with atmospheric liquid argon hosted by a cryostat as the one used in~\cite{bib:pdsp}. The dimensions of the fiducial volume are $12.0 \times 14.0 \times 58.2~\text{m}^3$, divided into four 3.5~m long drift volumes, as presented in Fig.~\ref{fig:detector}. The clearances between the detector and the cryostat are 0.5 m in the drift direction, 1 m in the vertical axis, and 1.9 m in the longitudinal axis. The electric field applied to the drift volumes is 500~V/cm. The collection planes are readout by contiguous Q-Pix pixels, supported by a light mechanical structure. This design ensures minimal engineered structure and materials.  In order to study the impact of light detection, we assume that the entire anode plane is sensitive to photons, which corresponds to a photon coverage of 37\% and we assume a 15\% quantum efficiency to reflect the efficiency of currently available commercial solutions for photon detection at 128 nm ~\cite{bib:hpk_vuv}. While the light detection system is very ambitious, it allows for a better understanding of how photon detection could help in reducing backgrounds. It can also serve as a benchmark to study innovative solutions for light readouts that could offer high detection coverage if the pixels were sensitive to both charge and light \cite{bib:ase}.


\textit{Low-background scenario.} For the low-background scenario, we follow the design proposed in the SLoMo proposal \cite{bib:slomo}. The detector is akin to the high background scenario, i.e. same readout systems and active volume but, in this case, filled with underground argon (UAr). An additional neutron absorber (either a water shield or a borated cryostat) is assumed to reduce the neutron contributions by 4 orders of magnitude and the radon background is suppressed by 3 orders of magnitude with dedicated argon re-circulation and purification systems.

\subsection{Cavern}
\noindent Neutrino detectors need to be located underground to be significantly shielded from backgrounds resulting from cosmic rays. Despite the suppressed cosmic-ray rates underground, several backgrounds remain and depend on the characteristics of the cavern where the detector is located, such as the rock composition.  For both the high and low-background scenarios, we consider a cavern where the 
 average composition of the rock is taken from Ref.~\cite{bib:capocci}.

\section{The Q-Pix readout }\label{sec:Q-PixOverview}


The Q-Pix scheme employs zero suppression, self-triggering pixels and dynamically established data networks to capture ionization signals in kiloton underground LArTPCs~\cite{Q-Pix_org}. The Q-Pix approach ensures high-precision sampling of low-energy depositions while remaining idle when nothing of interest occurs, embodying the electronic principle of ``Least Action". The readout remains in a low-power quiescent state when no ionization charge is present, effectively keeping the pixel ``OFF." However when signal is present, the pixel is ready to collect charge and quickly transition to an ``ON" state while maintaining low-power consumption to minimize heat dissipation in the liquid argon environment. 

\begin{figure}
    \centering
\includegraphics[width=0.7\textwidth]{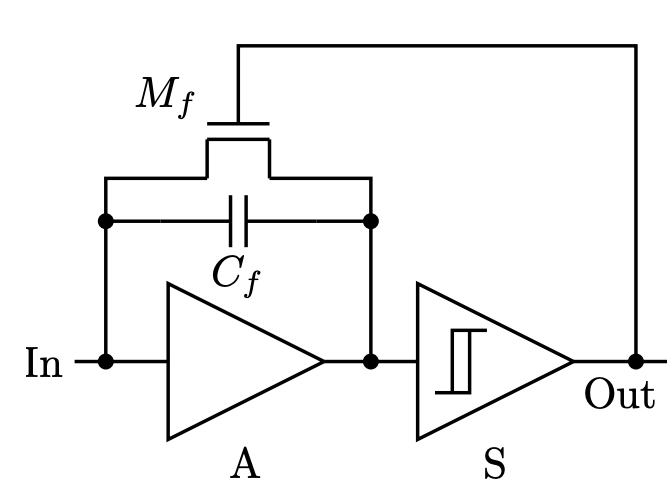}
    \caption{Illustration of the Q-Pix readout principle. When the accumulated charges in $C_f$, received by the input, meets the threshold, a Schmitt Trigger fires, which drains the charges in $C_f$ to reset the system. Figures taken from \cite{Q-Pix_SN}.}
    \label{fig:q_pix_reset}
\end{figure}

The concept is illustrated by the circuit block with a time-stamping mechanism shown in Fig.~\ref{fig:q_pix_reset}, where the charge-receiving pixel, labeled as ``Input'', continuously integrates the signal current with a Charge Sensitive Amplifier coupled with a feedback capacitor $C_f$. Once the integrated charge reaches a predetermined threshold $\Delta Q$, a Schmitt Trigger is fired, creating a reset of the system. The reset signal from the Schmitt Trigger will drain the charges accumulated in $C_f$ and triggers the recording of a time stamp of the local clock. Consecutive timestamps allow to calculate $\Delta t$, the time taken to integrate the unit of charge $\Delta Q$ on the capacitor.  $\Delta Q$ is defined by the threshold of the Schmitt Trigger, and the input current profile can be reconstructed entirely from the measured $\Delta t$, without having to record the entire waveform~\cite{Q-Pix_org, Q-Pix_COTS}. Since only 8-bits timestamps are recorded, this design uses a low-power architecture to address the high data rates associated with a high-granularity readout.

\subsection{Event simulation with Q-Pix }\label{sec:Q-PixSimulation}
\noindent In order to perform detailed physics studies, the Q-Pix consortium developed two main simulation packages:

\begin{enumerate}
    \item \texttt{QPixG4}: This software relies on Geant4 v4.11 \cite{AGOSTINELLI2003250} with the physics library FTFP\_BERT\_HP, combined with the low energy event generator MARLEY \cite{Marley_1} and the neutrino interaction generator GENIE \cite{GENIE} that models neutrino interactions in the detector environment.
    \item \texttt{QPixRTD}: This standalone code transforms the Geant4 output hits into Q-Pix resets, simulating the detector response.
\end{enumerate}

For simulating solar neutrinos, we use the MARLEY event generator \cite{Marley_1}, which generates low-energy CC or ES neutrino events from a given neutrino spectrum. The events are then run with the \texttt{QPixG4} package. In this study, the detector dimensions were set to a single $2.3 \times 6.0~ \text{m}^2$ anode sheet, corresponding to an active volume of $2.3 \times 6.0 \times 3.5~\text{m}^3$. Results from this sample were then scaled to the entire $12.0 \times 14.0 \times 58.2~\text{m}^3$ fiducial volume.

The \texttt{QPixRTD} package transforms Geant4 hits into pixel resets. The reset threshold and pixel size are both tunable parameters of the simulation that could be optimized for a specific physics goal. The transformation of hits into resets is achieved by first assigning a pixel index to each hit, in both horizontal and vertical coordinates. For each pixel, the charge of each hit is then added up until the chosen reset threshold is reached. Once the threshold is reached, the charge accumulated in that pixel is reset to 0, and the timestamp is recorded. This process is repeated while all charge from all hits associated with a pixel have been accounted for. If, after iterating over all hits, the charge accumulated in a pixel is lower than the reset threshold determined, no reset is produced and, instead, the charge accumulated is passed on to the subsequent event. This is a continuous process, meaning Q-Pix is continuously integrating the charge, with no need for a defined readout time window used in wire-based TPCs. No dead time between pixel resets is simulated, since we expect it to be negligible in the Charge-Integrate Replenishment scheme. The simulation output is as simple as it can be: pixel indices and reset timestamps~\cite{Q-Pix_COTS}.

\texttt{QPixRTD} assumes an anode plane fully tiled in $4 \times 4~\text{mm}^2$ contiguous pixels. Detector effects can be simulated by applying parametric models of diffusion, recombination and electron life-time. For recombination, \texttt{QPixRTD} uses the ``modified box model"~\cite{ArgoNeuT:2013kpa}. Diffusion is simulated using a Gaussian smearing with longitudinal and transverse diffusion constants respectively $D_L = 6.8223$~cm$^2$/s and $D_T = 13.1586$~cm$^2$/s \cite{LI2016160}. Electron life-time has negligible impact due to the assumption of high argon purity as achieved by other LArTPCs \cite{Meddage:2017lxo,Antonello:2014eha,Adams:2019wxc,bib:pdsp} and is, therefore, not considered.

\subsection{Event Reconstruction}\label{sec:Q-PixClustering}
\noindent  We define an event as the activity in the detector originating from a single initial physical process, i.e., a single solar neutrino or background decay. Each event may produce multiple resets across multiple pixels. Events are reconstructed by clustering Q-Pix resets.

The Q-Pix readout facilitates the implementation of clustering techniques by directly defining a 3D environment. First, electrons produced by ionization during an event are collected on the pixel plane, forming a set of data points arranged on a grid in the $x$ and $y$ spatial coordinates.
As the charge collected produces resets, the reset timestamps provide the time separation between resets, i.e., the temporal coordinate, which is later used with the electron drift velocity to obtain the $\Delta z$ spatial separation between resets. 

Once the raw reset coordinates are obtained, we use the density-based spatial clustering (DBSCAN) algorithm~\cite{ester1996density} to identify events of interest. The DBSCAN algorithm is a convenient choice as it only requires two parameters to determine clusters: the minimum number of resets, denoted as cluster threshold (CT), and the maximum separation ($\epsilon$) between the clusters.  Taking into account the different nature of the $x$ and $y$ coordinates and the temporal coordinate, we apply DBSCAN in two stages:
\begin{enumerate}
    \item The resets are clustered on the pixel plane. The only requirement is that the ``spatial" resets must be contiguous.
    \item The resets are clustered along their temporal coordinates. At this stage, $\epsilon$ and CT are varied so that we may choose optimal parameters for each study.
\end{enumerate} 
The maximum temporal separation between resets is fixed at $3$~$\mu$s, equivalent to $5$~mm. This interval is optimal for depositions in the MeV energy range~\cite{Q-Pix_SN}. 

\subsubsection{Energy reconstruction}
\label{sec:efficiency}

With the Q-Pix readout, the charge collected on the anode plane can be directly reconstructed from the number $N$ of resets produced, i.e., $\Qrec= N \times \Delta Q$, where $\Delta Q$ is the pixel charge threshold. However, since $\Delta Q$ is not infinitesimally small, some deposited charge $\Qdep$ may not be recorded in the correct event when $\Qdep$ is not an integer multiple of $\Delta Q$. That is, $\Qdep \geq N  \times\Delta Q$ and, therefore, $\Qdep \geq \Qrec$, resulting in $\Qrec$ being not entirely linear with $N$ and $\Qdep$. 
While non-triggered quanta could degrade the charge collection, their impact can be mitigated by energy calibration and by choosing a reset threshold $\Delta Q$ that is optimized for the different scientific goals. For low-energy neutrinos of the order of a few MeV, $\Delta Q = 1fC$ (6250 electrons) was found optimal~\cite{Q-Pix_SN}. 

The equivalent energy of a reset is obtained by multiplying the number of electrons in one $\Delta Q$ by the argon $W$ value of $W=23.6$~eV~\cite{TAKAHASHI1973123}. Therefore, in this study, $E_{\text{reset}}=23.6 \times 6250 \times 10^{-6} = 0.14$~MeV. If part of the charge released in an event is collected but does not fire the threshold trigger, it is kept as leftover charge for future events. This effect is expected to be small and can be corrected by a calibration procedure, and is thus not considered in this work. 

Another non-linear effect in the charge reconstruction is introduced by inaccuracies in the DBSCAN clustering. Resets belonging to the same event may not all be clustered together because the depositions are too far apart. In this case, smaller clusters may be formed, which will be rejected as the cluster threshold increases. Therefore, the threshold must be carefully selected to effectively discriminate background events while not eliminating legitimate signal clusters.

\section{Solar Neutrino Prediction}\label{sec:SolarPhysics}
\begin{figure}[htbp]
\centering
\includegraphics[width=0.9\textwidth]{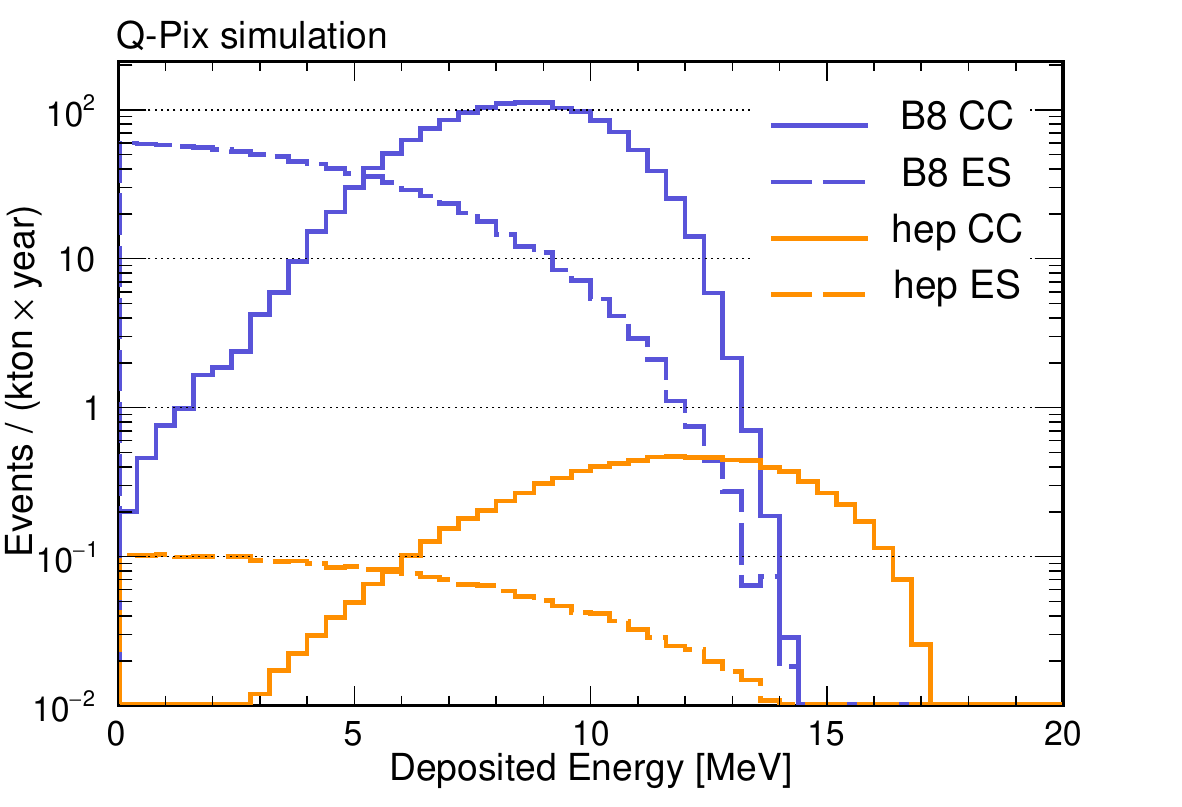}
\caption{\label{fig:signal} 
Energy spectra of the total expected B8 and hep electron neutrinos from CC and ES interactions in a 10 kiloton LAr active volume per year.}
\end{figure}

\noindent In this study, we target the detection of solar neutrinos produced either by the Boron-8 (B8) process or the proton-proton IV (hep) process interacting in the detector via either Charged Current (CC) interaction or Elastic Scattering (ES) interaction. These two specific solar neutrino processes were selected as the B8 process has the highest flux at energies above 2 MeV and the hep process produces the highest neutrino energies. Hep neutrinos have never been observed, but upper limits on their flux have been reported \cite{SNO:2020gqd,Borexino:2017uhp}. Table \ref{tab:signal_rates} and Fig. \ref{fig:signal} show the expected solar neutrino events in a 10~kiloton-year exposure for all four interaction categories (B8CC, B8ES, hepCC, and hepES). These predictions results from the solar neutrino flux from Ref.~\cite{Bahcall_2005} and the cross section of neutrino interactions in argon~\cite{Marley_1}.

\begin{table}[ht]
    \centering
    \setlength{\tabcolsep}{10pt}%
    \def\arraystretch{1.2}%
    \resizebox{1\textwidth}{!}{%
    \begin{tabular}{cc}
        \toprule
        Interaction Channel                                       & Events / (10 kton $\times$ year)      \\
        \midrule
        B8 CC ($\nu_e + \ce{^{40}Ar} \rightarrow e^- + \ce{^{40}K^*}$) & 14380 \\
        B8 ES ($\nu_e + e^- \rightarrow \nu_e + e^-$) & 9160 \\
        hep CC ($\nu_e + \ce{^{40}Ar} \rightarrow e^- + \ce{^{40}K^*}$)  & 86 \\
        hep ES ($\nu_e + e^- \rightarrow \nu_e + e^-$ )                  & 23 \\
        \bottomrule
    \end{tabular}
    }
    \caption{Expected total number of electron-neutrino interactions from the B8 and hep processes for charged-current (CC) and electron elastic scattering (ES) in 10 kiloton LAr active volume per year.}
    \label{tab:signal_rates}
\end{table}

\section{Background predictions}\label{sec:Backgrounds}

\noindent 

\noindent  Past solar experiments such as SNO \cite{bib:sno_final} and Borexino \cite{Borexino_2023,bib:borexino_cno_final} have shown that with a good control and understanding of backgrounds, detailed physics analyzes can be performed on solar neutrinos.

In this study, we construct a conservative background model representing a generic underground cavern environment. Our approach emphasizes worst-case scenarios by including all potential background sources that could compromise solar neutrino sensitivity below 20 MeV. While the model is not tailored to a specific site, it is designed to capture a full range of rare backgrounds that could impact future measurements. We emphasize that the feasibility of solar neutrino detection is highly dependent on site-specific environmental conditions—including but not limited to rock composition, shielding design, and detector construction. Any experiment-specific implementation is beyond the scope of this work.
Finally, we note that dominant backgrounds in the high energy tail (above 11 MeV) are severely under-constrained by current data. A central conclusion of this work is the urgent need for dedicated and site-specific essays to enable robust solar neutrino analyses in future large-scale detectors.


We classify backgrounds in two categories according to their origin. Internal backgrounds are those originating in the fiducial volume of the detector, either from radioactive decays in the liquid argon itself  (referred to as ``bulk argon radioactivity") or from radioactive isotopes contained in detector components. External backgrounds are those that originate outside of the detector and propagate into the fiducial volume, and are generated from decays in the cavern walls (including the rock concrete and shotcrete) and in the cryostat. External backgrounds are mainly neutrons and $\gamma$ rays. The mean free path of $\alpha$ and $\beta$ particles is too short to reach the liquid argon. The details of the backgrounds considered here are summarized in Table~\ref{tab:bgs}.  Unless it is mentioned otherwise in the following sub-sections, all backgrounds are simulated using QPixG4.

\subsection{Internal backgrounds}
\noindent In this section we describe the internal backgrounds that originate in the argon bulk or in the detector components.

\subsubsection{Argon radioactivity}
\noindent Commercially available argon commonly used in LArTPCs is obtained by liquefying argon from the atmosphere. In its natural isotopic composition, atmospheric argon consists of the stable \ce{^{40}Ar} isotope and trace amounts of the radioactive idotopes \ce{^{37}Ar} ($T_{1/2}=35$ d, $Q_{\beta}=0.813$ MeV), \ce{^{39}Ar} ($T_{1/2}=268$ y, $Q_{\beta}=0.565$ MeV), and \ce{^{42}Ar} ($T_{1/2}=32.9$ y, $Q_{\beta}=0.599$ MeV). Here, $T_{1/2}$ and $Q_{\beta}$ refer to the half-life and the energy decay components of the corresponding isotope, respectively. While \ce{^{37}Ar} decays via electron capture, \ce{^{39}Ar} and \ce{^{42}Ar} decay via $\beta$ emission producing an electron propagating in the detector. If their energy lies between 0.5 and 18.5 MeV, these $\beta$-electrons can mimic the signature of the solar neutrino events that also produce an electron. We do not consider the decay of \ce{^{37}Ar}, since it has a short lifetime (35 days) and we assume that all related activity will be negligible by the time of data taking. Activity of \ce{^{39}Ar} and \ce{^{42}Ar} in atmospheric LAr have been measured by the WARP~\cite{bib:ar39}, GERDA~\cite{bib:ar42_1}, DEAP-600 \cite{bib:ar42_2} and DBA~\cite{bib:ar42_3} experiments to be approximately 1~Bq/kg and 50-100~$\mu$Bq/kg, respectively. In this study we use 1~Bq/kg for \ce{^{39}Ar}, 100~$\mu$Bq/kg for \ce{^{42}Ar}.
We also consider \ce{^{85}Kr} ($T_{1/2}=10.7$~y, $Q_{\beta}=0.687$~MeV), which  
decays via$\beta$ emission. While the exact \ce{^{85}Kr} activity highly depends on the quality of the LAr batch and can vary by up to a factor three, here we assume an activity of 0.1 Bq/kg~\cite{bib:snowmass_lowenergy_lartpc,bib:ar39}.

The \ce{^{39}Ar} and \ce{^{85}Kr} daughters are stable. However, \ce{^{42}Ar} decays to \ce{^{42}K}, which is unstable. \ce{^{42}K} $\beta$ decays to the ground state of \ce{^{42}Ca} ($T_{1/2}=12$~h, $Q_{\beta}=3.525$~MeV) with 82\% branching ratio, and the remaining 18\% to an excited state of \ce{^{42}Ca}, emitting an additional 1.524 MeV $\gamma$ ray. If the \ce{^{42}K} daughter remains positively charged, this background chain could  potentially be reduced if the \ce{^{42}K^{+}} drifts to the cathode before it decays. However, there are no measurements of ion survival probability for \ce{^{42}K^{+}} in LAr. Furthermore, at these energies, the enormous background rate would be negligibly impacted by a potential reduction through ion drift. Thus, we make the conservative assumption that the activity of any nucleus daughter is the same as its parent.

The use of underground argon in the low-background scenario significantly reduces internal backgrounds, as discussed in the SLoMo proposal ~\cite{bib:slomo}. Since UAr is less activated by cosmic-ray interactions, the radioactive isotopes in UAr are suppressed. The assumed \ce{^{39}Ar} and \ce{^{85}Kr} activities in UAr is 0.73 mBq/kg and 2 mBq/kg, respectively, as measured by DarkSide~\cite{bib:darkside_UAr,bib:darkside_more_UAr} -- a reduction of three orders of magnitude in both cases. There are no measurements of \ce{^{42}Ar} activity in UAr, but in line with the other two isotopes, we assume a reduction of three orders of magnitude for the low-background scenario~\cite{bib:slomo, bib:ar42_uar}.

\subsubsection{Radon decay chains}
\label{sec:background_radon}
\noindent Radon emanates from any material that contains Uranium and Thorium isotopes, such as detector components and surrounding materials like rocks, concrete and shotcrete, and can diffuse inside the detector. Here, we treat radon isotopes and their daughters as an internal background since their activity will primarily occur inside the detector. In line with Ref.~\cite{bib:slomo}, we assume an activity of 1 mBq/kg for \ce{^{222}Rn} originating from the \ce{^{238}U} chain. Since the measured \ce{^{232}Th} abundance is 1.5 to 5 times higher than \ce{^{238}U} \cite{bib:capocci,bib:cavern_bg_1,bib:cavern_bg_2,bib:cavern_bg_3} and \ce{^{220}Rn} is a daughter of \ce{^{232}Th}, we assume a conservative \ce{^{220}Rn} activity of 5 mBq/kg, 5 times larger than \ce{^{222}Rn}. Finally, for \ce{^{219}Rn}, which originates from \ce{^{235}U}, we estimate an activity of 7 ~\textmu Bq/kg \cite{bib:ur} based on the relative abundance of uranium isotopes. Each of the above radon isotopes generates a radioactive decay chain that includes $\alpha$ and $\beta$ decays with different energies. The decays used in this study are presented in Table~\ref{tab:bgs}. The number of expected events for each decay of the chain is computed considering secular equilibrium, where the rate of the daughter is the parent decay rate multiplied by its branching ratio and by its decay probability in a one-year time window. For long-lived isotopes, such as \ce{^{210}Pb}, an accumulation and thus an activity increase is expected over time. However, as will be seen later, the results presented here are independent of these particular decays and any subsequent decay.

As previously mentioned, $\beta$ emitters and solar neutrinos both generate one primary electron propagating in the detector. Even though most of the background $\beta$-electrons have a maximum energy below $2$-$3$~MeV, thallium isotopes \ce{^{208}Tl} and \ce{^{210}Tl} have a higher $Q_{\beta}$ of $5$ and $5.4$~MeV, respectively.  The electrons can carry up to $2.38$~MeV for \ce{^{208}Tl} decays, with the remaining energy emitted as $\gamma$ rays, and up to $4.4$~MeV for \ce{^{210}Tl} decays. We estimate $\sim 10^{7}$ \ce{^{210}Tl} decays per 10 kton-year exposure, which is two orders of magnitude more than the total number of expected solar neutrino events with energies $<4.4$~MeV, making this a significant source of background.

\begin{figure}[htbp]
\centering
\includegraphics[width=0.9\textwidth]{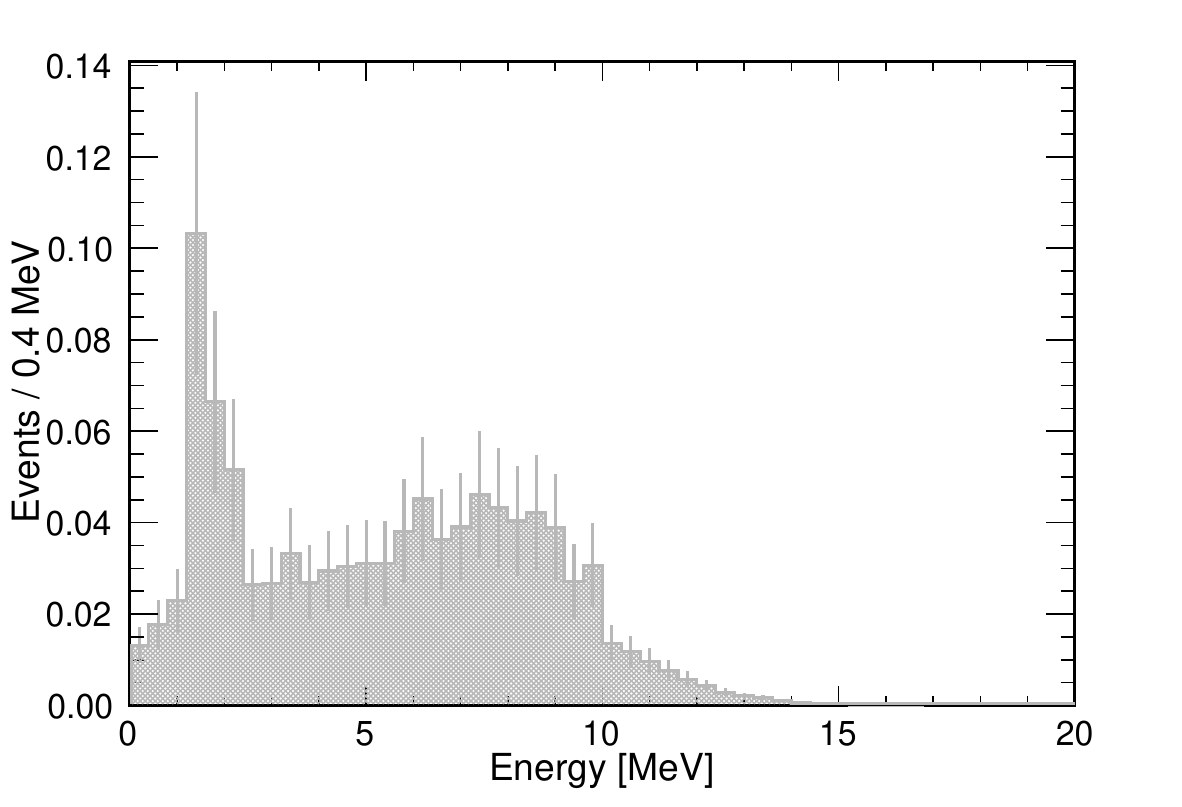}
\caption{\label{fig:alphacap}
Simulated energy spectrum of the $\gamma$ rays produced in \textalpha-capture processes in argon. The error bars represent the uncertainty on the cross-section measurement. The spectrum is normalized to 1. }
\end{figure}

While ionizing $\alpha$ particles from $\alpha$-emitters can be discriminated from solar neutrinos via their different ionization profiles, $\alpha$ particles can be captured by \ce{^{36}Ar}, \ce{^{38}Ar}, and \ce{^{40}Ar} in ``giant resonances" that de-excite and emit $\gamma$ rays with energies between 1 and $17$~MeV~\cite{bib:alpha_capture}. These $\gamma$ rays are an important background for solar neutrino detection, since — just like solar neutrinos — they can produce electrons in the active volume by Compton scattering or pair production. The $\alpha$-capture processes are not yet measured with good precision (currently there is a $30\%$ uncertainty on cross-section measurements \cite{bib:alpha_capture}). Nevertheless, we attempt to estimate the rates of this additional $\gamma$ background, given its potential impact on the study of solar neutrinos. The $\alpha$-capture differential cross section in argon is reported in~\cite{bib:alpha_capture} for $\alpha$ energies between 5.5 and 15 MeV, at 90$^{\circ}$ with respect to the beam direction. They are also reported to happen down to 3~MeV of $\alpha$ energies in~\cite{bib:alpha_capture_2}. Starting from those values and considering the isotropy of our use-case, we calculate the interaction probability for the different $\alpha$ energies under the assumption of continuous slowing down approximation down to 3~MeV, integrating over the full solid angle. We obtain an $\alpha$-capture rate of almost $\sim 10^{4}$ per 10~kiloton per year. We also simulate this process in QPixG4 by generating $\alpha$ particles isotropically inside the detector with the energies and relative frequencies presented in Table \ref{tab:bgs}, obtaining a rate of $\sim 10^{6}$ $\alpha$ captures per 10~kiloton per year, which seems to overestimate the measurements presented in Ref.~\cite{bib:alpha_capture}. 
The obtained $\gamma$ spectrum is presented in Figure \ref{fig:alphacap}. Given the profound impact of $\alpha$-capture backgrounds on the study of solar neutrinos, the large uncertainties on the measured process~\cite{bib:alpha_capture}, and the potential inaccuracy of simulations relying on minimal data available, we conservatively assume the rate predicted by Geant4 with the caveat that new measurements of $\alpha$ capture are of paramount importance to draw any final conclusion on the observability of hep neutrinos and the study of solar neutrinos in general.


Alpha particles can also interact with argon nuclei via \ce{^{40}Ar}(\textalpha,n\textgamma)\ce{^{43}Ca} \cite{43ca}, emitting neutrons and $\gamma$ rays in the final state. Based on the simulation, we expect $\sim 10^{7}$ of these reactions per year. The simulation of this process shows that the $\gamma$ has an energy range of 0.3-6.2 MeV, similar to what is reported in Ref.~\cite{bib:n43}, while neutron energy can reach 3.6 MeV. As these energies are below the energies of external neutrons and $\gamma$ rays presented in the following sections, and their rate is significantly lower, we neglect their contributions. 

In addition to the $\alpha$ particles produced by radon decay chains, one Rn daughter, \ce{^{210}Tl}, can also undergo $\beta$-delayed neutron emission. Whenever the $\beta$ decay populates excited states in the daughter nucleus above the neutron separation energy, neutron emission becomes energetically allowed. This neutron carries the excess energy of 5.189 MeV. As in the previous case, the rate of this particular decay is very low compared to that of external neutrons, and we therefore do not include it in our background model.

The summary of all the decays coming from radon isotopes, with their respective activities and half-lives, is presented in Table \ref{tab:bgs}. As previously mentioned, some nuclei resulting from radioactive decays may remain positively charged, thus drifting towards the cathode and potentially being removed if their half-lives are larger than the drift time. EXO-200 ~\cite{bib:ion_surv_LXe} reports the ion survival probability of \ce{^{218}Po^{+}} (result of an $\alpha$ decay) and \ce{^{214}Bi^{+}} (result of a $\beta$ decay) in liquid xenon to be approximately 
$50\%$ and $75\%$, respectively, and DarkSide ~\cite{bib:ion_surv_LAr} reports the ion survival probability of \ce{^{218}Po^{+}} in LAr to be 37\% -- significantly lower than in LXe. The survival probability ultimately depends on the recombination effect, which depends on the electric field applied in the active volume, but it also depends on the LAr purity (as ions can recombine with electrons from impurities) and LAr flow dynamics, as ion drift velocity is of the same order of magnitude as LAr flow velocity. While this effect could reduce backgrounds, we conservatively assume that the activity of any nucleus daughter is the same as its parent given that the potential reduction is small compared to the difference in expected signal and background. For example, it takes four decays to produce \ce{^{208}Tl} (\ce{^{220}Rn} $\rightarrow$ \ce{^{216}Po} $\rightarrow$ \ce{^{212}Pb} $\rightarrow$ \ce{^{212}Bi} $\rightarrow$ \ce{^{208}Tl}). Assuming a 50\% ion survival probability, and assuming that all surviving ions are collected before decaying, the \ce{^{208}Tl} rate would be reduced from $10^{12}$ to $\sim 10^{10}$, still five orders of magnitude more than the number of expected solar neutrino events.

In the low-background scenario, the use of a dedicated radon purification system combined with careful fabrication and installation procedures can reduce the radon activity in the detector by three orders of magnitude~\cite{bib:slomo}. Such reduction has already been demonstrated in dark matter experiments such as DarkSide\cite{bib:radon_reduction_darkside} and DEAP-600 \cite{bib:radon_reduction_deap}. Therefore, when considering the low-background scenario, we reduce all decays associated with \ce{Rn} activity by three orders of magnitude.

\subsubsection{Radioactivity from detector components}
\noindent The last sources of internal background we consider are the detector components. The different detector parts contain radiocontaminants such as \ce{^{238}U}, \ce{^{235}U}, \ce{^{232}Th}, \ce{^{60}Co} or \ce{^{40}K} that can produce background events. However, since it is expected that the support structures needed for a Q-Pix readout will be simple and lightweight, we neglect the potential radioactive decays from the detector components. 
Finally, while \ce{^{238}U}, \ce{^{235}U} and \ce{^{232}Th} lead to the aforementioned radon chains, \ce{^{60}Co} and \ce{^{40}K} are low-energy $\beta$ emitters (\textless 1.5 MeV) that do not affect the conclusions presented at the end of this study. 

\subsection{External backgrounds}
\noindent In this section we describe the backgrounds
that originate outside the cryostat.
\subsubsection{Neutrons}
\label{sec:background_neutrons}
\begin{figure}[htbp]
\centering
\includegraphics[width=0.9\textwidth]{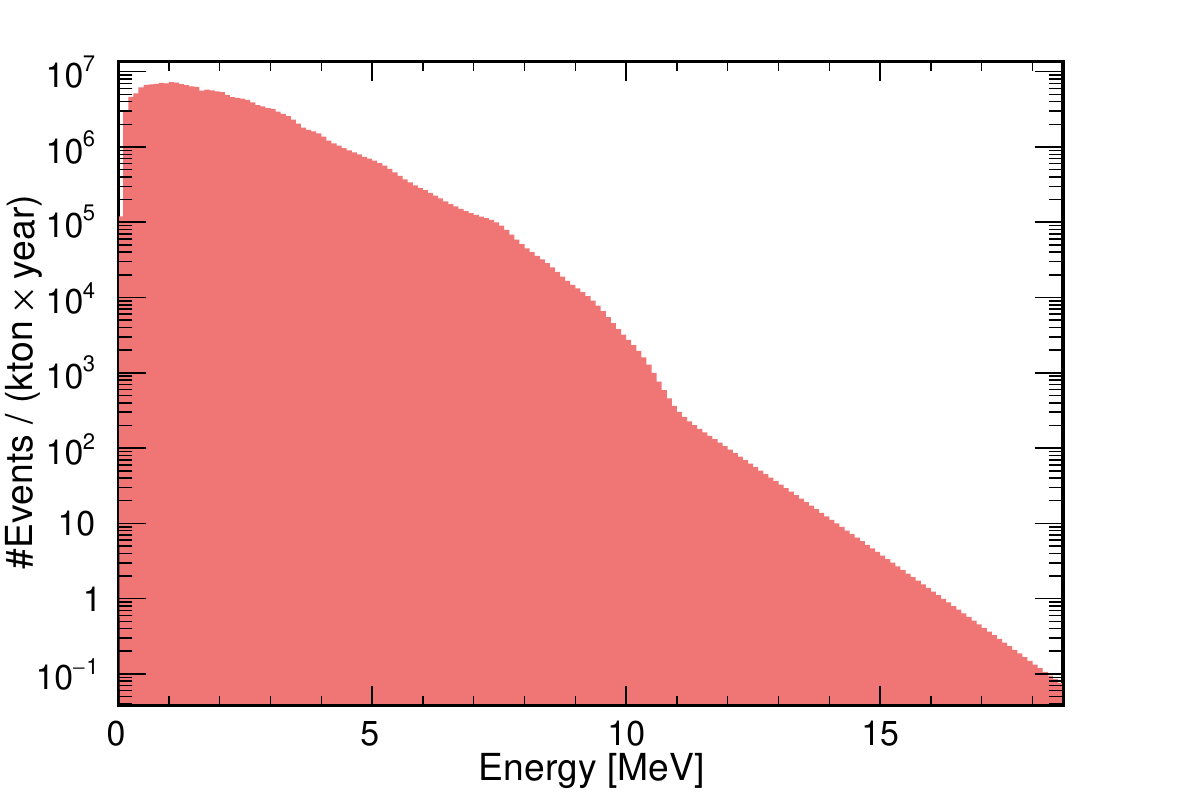}
\caption{\label{fig:neutrons} 
Energy spectrum for the simulated neutrons produced in the cavern walls and reaching the liquid-argon surface.}
\end{figure}

\noindent Neutrons are emitted in fission processes and $(\alpha, n)$ reactions created by $\alpha$ decays resulting from the \ce{^{232}Th} and \ce{^{238}U} chains. To estimate the neutron energy spectrum from the cavern, we use the material compositions in Ref.~\cite{bib:capocci}. The $(\alpha, n)$ components of the neutron spectrum used in this study are generated via the \texttt{NeuCBOT}~\cite{bib:neubot} software. The uranium spontaneous fission component of the neutron spectrum is calculated using the Watt spectrum, with a neutron yield of 2 neutrons per decay, similar to Ref.~\cite{bib:capocci}. The neutron flux in the cavern is assumed to be $1\cdot10^{-5}$ n/cm$^{2}$/s, in line with Ref.~\cite{bib:slomo}, which corresponds to neutrons penetrating the LAr surface with a frequency of 12 Hz. The total simulated neutron energy spectrum is shown in Fig.~\ref{fig:neutrons}. Neutrons are then propagated in LAr using the QPixG4 package.

In argon, neutrons with energy above 1.46~MeV can undergo inelastic scattering ~\cite{bib:delayed_gamma_2,b:n_inelastic}. This can generate $\gamma$ cascades with energies reported in the range of [1.46, 11.77]~MeV \cite{bib:delayed_gamma_2}, however the $\gamma$ rates decrease rapidly with energy. These $\gamma$ rays are an important background, as they can generate electrons by Compton scattering or pair production mimicking solar neutrino signals. Combining the measured neutron-argon inelastic cross section \cite{b:n_inelastic}, the assumed neutron flux, and our QPixG4 simulation, we estimate a neutron inelastic scattering frequency of 0.2 Hz per 10 kiloton.

Below 1.46 MeV, neutrons undergo elastic scattering with high rates in the keV energy region until they thermalize. Since this is far below our energy range of interest in the MeV, we neglect this contribution. Finally, thermal neutrons can be captured by argon nuclei. Neutron captures on \ce{^{40}Ar} and \ce{^{36}Ar} generate $\gamma$ cascades up to 6.1~MeV and 8.8~MeV, respectively~\cite{bib:n41,bib:n37}. Neutron captures on impurities as \ce{^{14}N} can reach higher energies~\cite{bib:14n_ncapture}, but given the expected purity levels of LAr~\cite{bib:pdsp}, we consider these contributions negligible. The rate of neutron captures in the active volume of the detector is found to be 1~Hz per 10 kton. To correctly simulate the $\gamma$ cascades, we re-weighted the QPixG4 output for the $\gamma$ cascade process to match the measured data in Ref~\cite{bib:neutron_resonances}.

In the low-background scenario, a water shield of 50 cm in combination with a borated cryostat reduce the neutron backgrounds by four orders of magnitude \cite{bib:slomo}. We therefore reduce the simulated neutron rates shown in Fig.~\ref{fig:neutrons} by $10^{4}$ for the low-background scenario.

\subsubsection{$\gamma$ rays}
\label{sec:background_gammas}
\noindent The interactions of $\gamma$ rays in the liquid argon are an irreducible background as they result in the production of single electrons via Compton scattering or pair production. It is therefore crucial to estimate the rates of $\gamma$ rays produced in the detector surroundings, referred to here as ``external $\gamma$ rays". 
The $\gamma$ rays from radioactive decays can reach energies of approximately $3$~MeV. Neutrons produced from fission or (\textalpha,n) reactions can be captured in the cavern, in the concrete or shootcrete of the cavern walls, and in the cryostat, generating more energetic $\gamma$ rays, with energies up to 11.5~MeV~\cite{bib:gamma_cascade}. Finally, $\alpha$ particles generated in the aforementioned radioactive chains and fission reactions can interact or be captured by nuclei in the cavern and the shotcrete, generating even more energetic $\gamma$ rays. For example, $\alpha$-captures in magnesium and silicon lead to de-excitation energies of more than $17$~MeV~\cite{bib:alphacapture_mg}. The effects of $\gamma$ rays are minimally mitigated by the cryostat material. They are expected to propagate of the order of $10$--$100$~cm in liquid argon. 

\begin{figure}[htbp]
\centering
\includegraphics[width=0.9\textwidth]{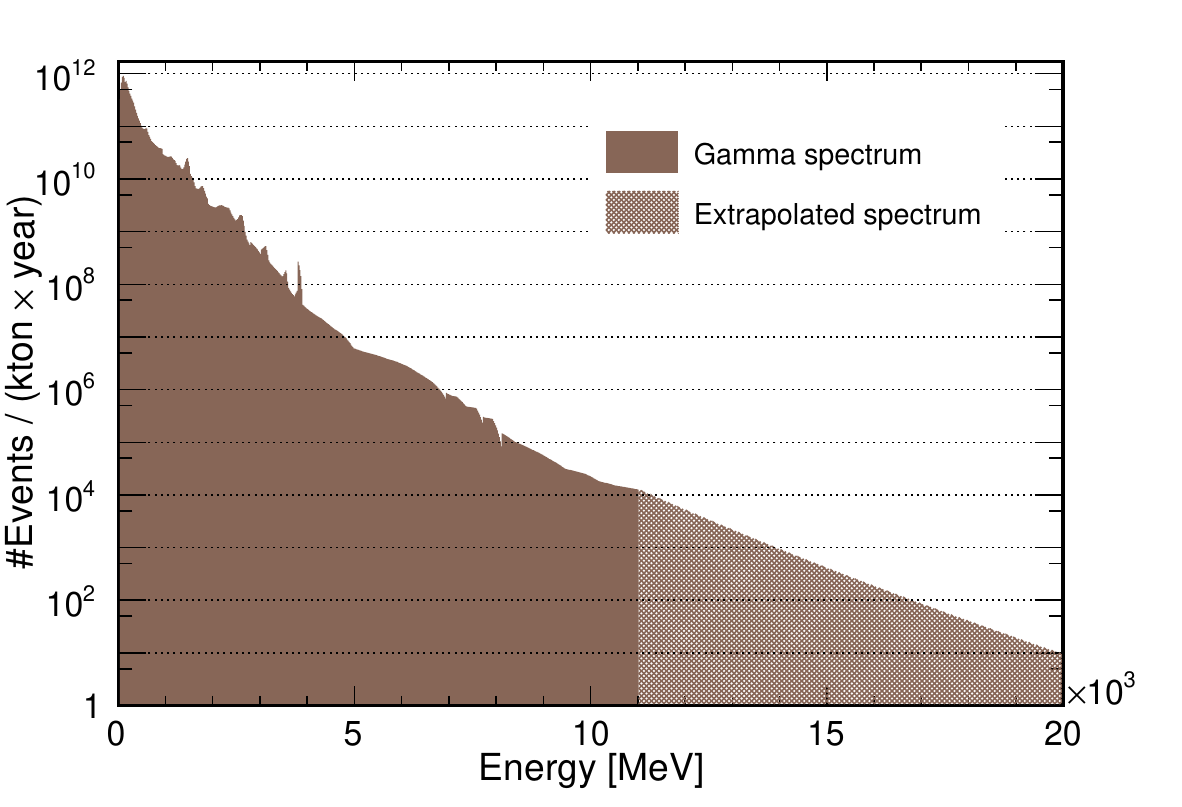}
\caption{\label{fig:gammas} 
True energy spectrum of the external $\gamma$ rays incident on the LAr interfaces based on measurements performed in underground facilities around the world~\cite{bib:LZ,bib:LNGS_gamma,bib:sno_gamma,bib:amore_gamma,bib:jinping_gamma,bib:sk_gamma}. The cross-hatched tail corresponds to the exponential extrapolation.}
\end{figure}

To quantify the impact of external $\gamma$ rays on our conclusions, we perform a detailed review of $\gamma$-ray rates as measured in existing experiments to validate simulations. Even though the composition of the cavern rocks and building materials (e.g., shotcrete) affects the exact $\gamma$ flux of all underground laboratories, the observed similarities in the relative frequencies of the $\gamma$-ray energies build confidence in the spectrum considered in this work. As our starting point, we use the $\gamma$-ray spectrum measured up to $3$~MeV by the LZ Collaboration in the Davis cavern~\cite{bib:LZ}. They report a total $\gamma$-ray flux of 1.9~$\gamma$/cm$^{2}$/s, which we take as normalization. This implies   that $\gamma$ rays enter the liquid-argon surface with a rate of $4$~MHz. To expand this spectrum to higher $\gamma$ energies, we leverage the measurements performed in other underground facilities such as Gran Sasso~\cite{bib:LNGS_gamma}, SNOLAB~\cite{bib:sno_gamma}, YangYang~\cite{bib:amore_gamma}, Jinping~\cite{bib:jinping_gamma} and Kamiokande~\cite{bib:sk_gamma}. We generate a weighted average spectrum that is added to the one reported by the LZ Collaboration. This new spectrum reaches to energies of $\approx 11$~MeV. As discussed previously, physical processes can occur in the cavern that generate even more energetic photons. To account for these, we arbitrarily choose to extrapolate the averaged spectrum up to $20$~MeV exponentially. The resulting spectrum is presented in Fig.\ref{fig:gammas}. In the spectra reported in underground facilities around the world, the fraction of $\gamma$ rays above 5 MeV covers a wide range of $10^{-8}$-- $10^{-5}$. This variability in the estimate of the high energy component is relevant as high rates of $\gamma$ backgrounds above 5 MeV can dominate in the solar neutrino signal region for B8 and hep neutrinos. Hence, dedicated measurements of the external $\gamma$ spectrum, especially at energies above 5 MeV, will be crucial to estimate sensitivities to solar neutrinos.

External $\gamma$ background is not considered in the SLoMo proposal~\cite{bib:slomo}. We therefore do not assume further reduction strategies for the $\gamma$ background in our low-background scenario and use the predicted rates discussed in this section.

\begin{table*}[htb]
    \centering
    \setlength{\tabcolsep}{10pt}
    \def\arraystretch{1.15}%
    \resizebox{1\textwidth}{!}{%
    \begin{tabular}{|c|cccccc|cc|c|c|}
        \toprule
        Isotope      & Half-life (s)     & Decay   & $Q$ (MeV) & Daughter & BR ($\%$) & Rate (Bq/kg) & 
\multicolumn{2}{c|}{Expected rate for 10~kt$\cdot$yr}     & Ref. \\
 &  & & & & &  & High Bkg. & Low Bkg.& \\
        \midrule
        \ce{^{39}Ar}  & 8.5$\times10^{9}$  & $\beta$ & 0.565     & \ce{^{39}K}      & 100      & 1.1                 & $\sim10^{14}$ & $\sim10^{11}$  & \multirow{4}{1cm}{\makecell{\cite{bib:slomo}\\\cite{bib:ar39,bib:ar42_1,bib:ar42_2,bib:ar42_3,bib:snowmass_lowenergy_lartpc,bib:darkside_UAr,bib:darkside_more_UAr,bib:ar42_uar}}} \\
        \cline{1-9} 
        \ce{^{42}Ar}  & 1.0$\times10^{9}$  & $\beta$ & 0.599     & \ce{^{42}K}      & 100      & 1.0$\times10^{-4}$  & $\sim10^{10}$ & $\sim10^{7}$   &                  \\
        \ce{^{42}K}   & 4.5$\times10^{4}$  & $\beta$ & 3.525    & \ce{^{42}Ca}     & 100      & 1.0$\times10^{-4}$  & $\sim10^{10}$ & $\sim10^{7}$   &                 \\
        \cline{1-9}
        \ce{^{85}Kr}  & 3.9$\times10^{4}$  & $\beta$ & 0.687     & \ce{^{85}Rb}     & 100      & 1.0$\times10^{-1}$  & $\sim10^{13}$ & $\sim10^{10}$  &                   \\
        \hline
        \hline
        \ce{^{219}Rn} & 4.0                & $\alpha$ & 6.946    & \ce{^{215}Po}    & 100      & 7.0$\times10^{-6}$  & $\sim10^{9}$  & $\sim10^{6}$ & \multirow{38}{1cm}{\makecell{\cite{bib:slomo,bib:capocci}\\\cite{bib:cavern_bg_1,bib:cavern_bg_2,bib:cavern_bg_3,bib:ur}\\\cite{bib:ion_surv_LXe,bib:ion_surv_LAr,bib:radon_reduction_darkside,bib:radon_reduction_deap}}} \\
        \ce{^{215}Po} & 1.8$\times10^{-3}$ & $\alpha$ & 7.526    & \ce{^{211}Pb}    & 99.99977 & 7.0$\times10^{-6}$  & $\sim10^{9}$  & $\sim10^{6}$   &                  \\
        \ce{^{215}Po} & 1.8$\times10^{-3}$ & $\beta$  & 0.721    & \ce{^{215}At}    & 0.00023  & 1.6$\times10^{-11}$ & $\sim10^{3}$  & $\sim10^{0}$   &                  \\
        \ce{^{211}Pb} & 0.6                & $\beta$ & 1.367    & \ce{^{211}Bi}    & 100      & 7.0$\times10^{-6}$  & $\sim10^{9}$  & $\sim10^{6}$   &                  \\
        \ce{^{215}At} & 1$\times10^{-4}$   & $\alpha$ & 8.178    & \ce{^{211}Bi}    & 100      & 1.6$\times10^{-11}$ & $\sim10^{3}$  & $\sim10^{0}$   &                  \\
        \ce{^{211}Bi} & 3.6$\times10^{-2}$ & $\alpha$ & 6.750    & \ce{^{207}Tl}    & 99.72    & 7.0$\times10^{-6}$  & $\sim10^{9}$  & $\sim10^{6}$   &                  \\
        \ce{^{211}Bi} & 3.6$\times10^{-2}$ & $\beta$  & 0.574    & \ce{^{211}Po}    & 0.28     & 2.0$\times10^{-8}$  & $\sim10^{6}$  & $\sim10^{3}$   &                  \\
        \ce{^{207}Tl} & 7.9$\times10^{-2}$ & $\beta$  & 1.418    & \ce{^{207}Pb}    & 100      & 7.0$\times10^{-6}$  & $\sim10^{9}$  & $\sim10^{6}$   &                  \\
        \ce{^{211}Po} & 0.5                & $\alpha$ & 7.594    & \ce{^{207}Pb}    & 100      & 2.0$\times10^{-8}$  & $\sim10^{6}$  & $\sim10^{3}$   &                  \\
        \cline{1-9}
        \ce{^{220}Rn} & 55.6               & $\alpha$ & 6.404    & \ce{^{216}Po}    & 100      & 5$\times10^{-3}$    & $\sim10^{12}$ & $\sim10^{9}$   &                  \\
        \ce{^{216}Po} & 0.1                & $\alpha$ & 6.906    & \ce{^{212}Pb}    & 100      & 5$\times10^{-3}$    & $\sim10^{12}$ & $\sim10^{9}$   &                  \\
        \ce{^{212}Pb} & 3.8$\times10^{4}$  & $\beta$  & 0.569    & \ce{^{212}Bi}    & 100      & 5$\times10^{-3}$    & $\sim10^{12}$ & $\sim10^{9}$   &                  \\
        \ce{^{212}Bi} & 3.6$\times10^{3}$  & $\beta$  & 2.251    & \ce{^{212}Po}    & 64       & 3.2$\times10^{-3}$  & $\sim10^{12}$ & $\sim10^{9}$   &                  \\
        \ce{^{212}Bi} & 3.6$\times10^{3}$  & $\alpha$ & 6.207    & \ce{^{208}Tl}    & 36       & 1.8$\times10^{-3}$  & $\sim10^{11}$ & $\sim10^{8}$   &                  \\
        \ce{^{208}Tl} & 1.8$\times10^{2}$  & $\beta$  & 4.999    & \ce{^{208}Pb}    & 100      & 3.2$\times10^{-3}$  & $\sim10^{11}$ & $\sim10^{8}$   &                  \\
        \ce{^{212}Po} & 3.0$\times10^{-7}$ & $\alpha$ & 8.954    & \ce{^{208}Pb}    & 100      & 1.8$\times10^{-3}$  & $\sim10^{12}$ & $\sim10^{9}$   &                  \\
        \cline{1-9}
        \ce{^{222}Rn} & 3.3$\times10^{5}$  & $\alpha$       & 5.590    & \ce{^{218}Po}    & 100       & 1.0$\times10^{-3}$   & $\sim10^{11}$ & $\sim10^{8}$ &                    \\
        \ce{^{218}Po} & 1.9$\times10^{2}$  & $\alpha$       & 6.114    & \ce{^{214}Pb}    & 99.98     & 1.0$\times10^{-3}$   & $\sim10^{11}$ & $\sim10^{8}$ &                    \\
        \ce{^{218}Po} & 1.9$\times10^{2}$  & $\beta$        & 0.259    & \ce{^{218}At}    & 0.02      & 2.0$\times10^{-7}$   & $\sim10^{7}$  & $\sim10^{4}$  &                   \\
        \ce{^{214}Pb} & 1.6$\times10^{3}$  & $\beta$        & 1.018    & \ce{^{214}Bi}    & 100       & 1.0$\times10^{-3}$   & $\sim10^{11}$ & $\sim10^{8}$  &                   \\
        \ce{^{218}At} & 1.3                & $\alpha$       & 6.874    & \ce{^{214}Bi}    & 99.9      & 2.0$\times10^{-7}$   & $\sim10^{7}$  & $\sim10^{4}$  &                   \\
        \ce{^{218}At} & 1.3                & $\beta$        & 2.881    & \ce{^{218}Rn}    & 0.1       & 2.0$\times10^{-10}$  & $\sim10^{4}$  & $\sim10^{1}$  &                   \\
        \ce{^{218}Rn} & 2.1$\times10^{3}$  & $\alpha$       & 7.262    & \ce{^{214}Po}    & 100       & 2.0$\times10^{-10}$  & $\sim10^{4}$  & $\sim10^{1}$  &                   \\
        \ce{^{214}Bi} & 1.2$\times10^{3}$  & $\beta$        & 3.269    & \ce{^{214}Po}    & 99.97     & 1.0$\times10^{-3}$   & $\sim10^{11}$ & $\sim10^{8}$  &                   \\
        \ce{^{214}Bi} & 1.2$\times10^{3}$  & $\alpha$       & 5.621    & \ce{^{210}Tl}    & 0.021     & 2.1$\times10^{-7}$   & $\sim10^{7}$  & $\sim10^{4}$  &                   \\
        \ce{^{214}Bi} & 1.2$\times10^{3}$  & $\alpha/\beta$ & 3.270    & \ce{^{210}Pb}    & 0.003     & 3.0$\times10^{-8}$   & $\sim10^{6}$  & $\sim10^{3}$  &                   \\
        \ce{^{214}Po} & 1.6$\times10^{-4}$ & $\alpha$       & 7.883    & \ce{^{210}Pb}    & 100       & 1.0$\times10^{-3}$   & $\sim10^{11}$ & $\sim10^{8}$  &                   \\
        \ce{^{210}Tl} & 78.0               & $\beta$        & 5.483    & \ce{^{210}Pb}    & 99.991    & 2.1$\times10^{-7}$   & $\sim10^{7}$ & $\sim10^{4}$   &                  \\
        \ce{^{210}Tl} & 78.0               & $\beta/n$      & 0.294    & \ce{^{209}Pb}    & 0.009     & 1.9$\times10^{-11}$  & $\sim10^{3}$ & $\sim10^{0}$   &                  \\
        \ce{^{209}Pb} & 1.2$\times10^{4}$  & $\beta$        & 0.644    & \ce{^{209}Bi}    & 100       & 1.9$\times10^{-11}$  & $\sim10^{3}$ & $\sim10^{0}$   &                  \\
        \ce{^{209}Bi} & 6.3$\times10^{26}$ & $\alpha$       & 3.137    & \ce{^{205}Tl}    & 100       & 0.0                  & $\sim10^{0}$ & $\sim10^{0}$   &                  \\
        \ce{^{210}Pb} & 7.0$\times10^{8}$  & $\beta$        & 0.063    & \ce{^{210}Bi}    & 99.999998 & 3.1$\times10^{-5}$   & $\sim10^{9}$ & $\sim10^{6}$   &                  \\
        \ce{^{210}Pb} & 7.0$\times10^{8}$  & $\alpha$       & 3.792    & \ce{^{206}Hg}    & 0.000002  & 5.8$\times10^{-13}$  & $\sim10^{2}$ & $\sim10^{-1}$  &                   \\
        \ce{^{206}Hg} & 4.9$\times10^{2}$  & $\beta$        & 1.308    & \ce{^{206}Tl}    & 100       & 5.8$\times10^{-13}$  & $\sim10^{2}$ & $\sim10^{-1}$  &                   \\
        \ce{^{210}Bi} & 4.3$\times10^{5}$ & $\alpha$        & 5.036    & \ce{^{206}Tl}    & 0.0001    & 3.1$\times10^{-11}$  & $\sim10^{3}$ & $\sim10^{0}$   &                  \\
        \ce{^{210}Bi} & 4.3$\times10^{5}$ & $\beta$         & 1.162    & \ce{^{210}Po}    & 99.9999   & 3.1$\times10^{-5}$   & $\sim10^{9}$ & $\sim10^{6}$   &                  \\
        \ce{^{206}Tl} & 2.5$\times10^{2}$ & $\beta$         & 1.532    & \ce{^{206}Pb}    & 100       & 5.8$\times10^{-13}$  & $\sim10^{2}$ & $\sim10^{-1}$  &                   \\
        \ce{^{210}Po} & 1.2$\times10^{7}$ & $\alpha$        & 5.407    & \ce{^{206}Pb}    & 100       & 2.6$\times10^{-5}$   & $\sim10^{9}$ & $\sim10^{6}$   &                  \\
        \hline
        $\alpha$-capture  &                   &                &          &                  &         &                   & $\sim10^{6}$  & $\sim10^{3}$ & \cite{bib:alpha_capture,bib:n43} \\
        \hline
        \hline
        n-inelastic &                   &                &          &                  &         &                   & $\sim10^{6}$  & $\sim10^{2}$ & \multirow[b]{2}{1cm}[2.75mm]{\makecell{\cite{bib:slomo,bib:capocci}\\\cite{bib:neubot,bib:delayed_gamma_2,b:n_inelastic,b:n_inelastic,bib:n41,bib:n37,bib:14n_ncapture,bib:neutron_resonances}}} \\
        n-capture   &                   &                &          &                  &         &                   & $\sim10^{7}$  & $\sim10^{3}$ &                    \\
        \hline
        Cavern $\gamma$     &                   &                &          &                  &         &                   & $\sim10^{11}$ & $\sim10^{11}$ & \cite{bib:LZ,bib:LNGS_gamma,bib:sno_gamma,bib:amore_gamma,bib:jinping_gamma,bib:sk_gamma} \\
        \bottomrule
    \end{tabular}
    }
    \caption{Summary of all the background processes considered for the study discussed in Section ~\ref{sec:Backgrounds}. All radioactive chains are separated by a solid line and they stop when a stable daughter is reached. The number of expected events for each decay of the chain is computed considering secular equilibrium, where the rate of the daughter is the product of the parent decay rate, its branching ratio, and its decay probability within a one-year time window.}
   \label{tab:bgs}
\end{table*}

\section{Results}\label{sec:Results}
\noindent In this section, we present the results obtained from our simulations of solar neutrinos and backgrounds using the assumptions described in Sections \ref{sec:SolarPhysics} and \ref{sec:Backgrounds}. We investigate several potential background reduction strategies and we report the data rates expected in the Q-Pix data acquisition that would allow for a continuous readout, essential to study solar neutrinos.

\begin{figure}[h!]
\centering
\includegraphics[width=\textwidth]{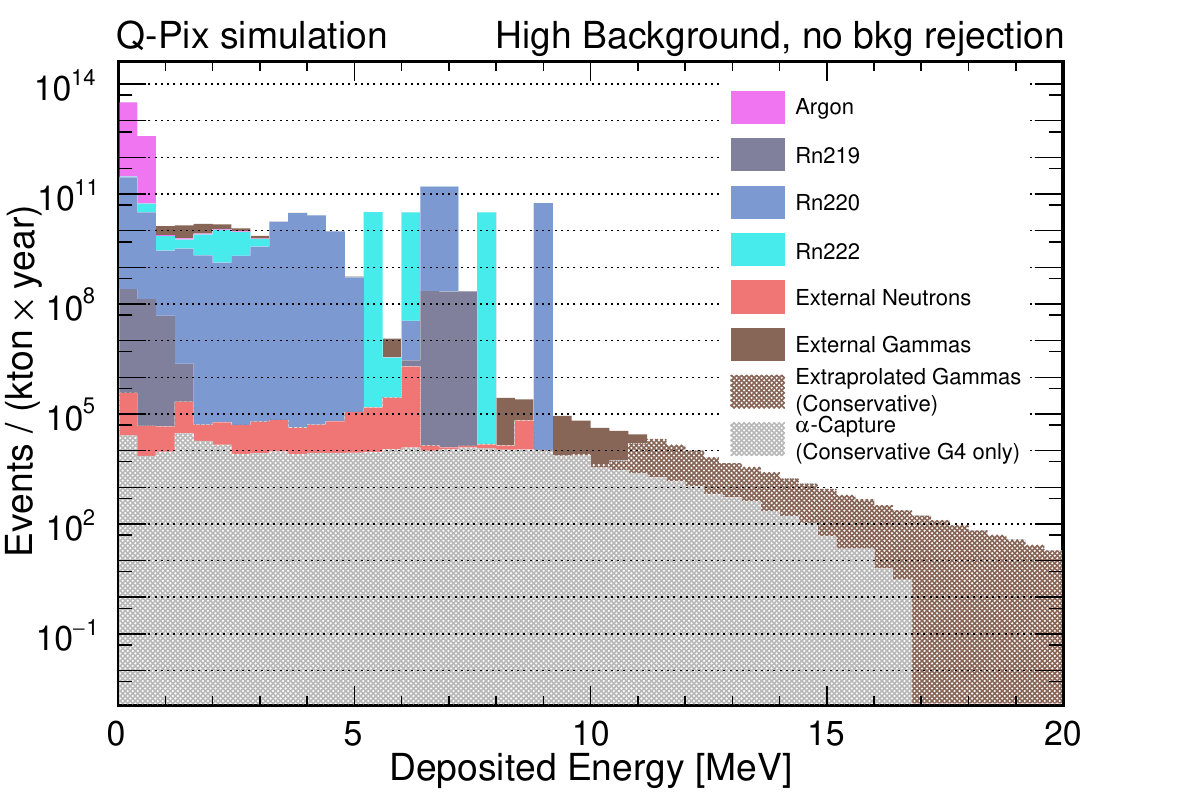}
\includegraphics[width=\textwidth]{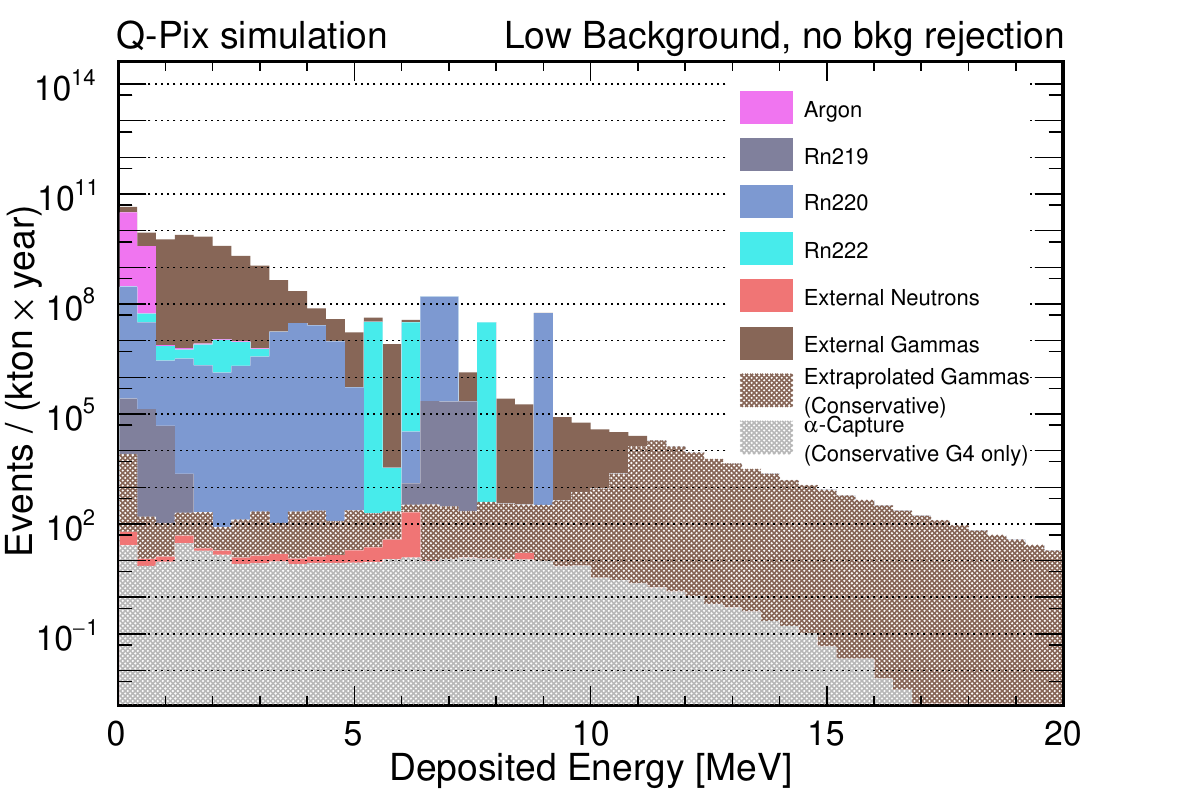}
\caption{\label{fig:spectrum}Energy spectra of all signal and background sources for the high-background scenario (top) and low-background scenario (bottom) before any selection tool has been applied.}
\end{figure}

Figure~\ref{fig:spectrum}a shows the spectra of deposited energy for 
all the backgrounds considered (see Section~\ref{sec:Backgrounds}) in the high background scenario. Even in the low background scenario, the spectrum of solar neutrinos is at least three orders of magnitude lower than the background when no background rejection strategy is applied. Studying solar neutrinos below 8 MeV is virtually impossible, as the background rates from radioactive decays, which are the ones predicted with the highest confidence in this study, are between 8 and 11 orders of magnitude above the solar neutrino signal. Above an energy of 8 MeV, the $\gamma$-ray backgrounds (both from external $\gamma$ rays and from $\alpha$-captures) are the dominant ones. In the 8-12 MeV window, the detection of solar neutrinos appear possible, but will strongly rely on experimentally constraining the background precisely.

Figure~\ref{fig:spectrum} shows the same deposited energy spectra for the low-background detector scenario. While many backgrounds are significantly reduced, the cavern $\gamma$ rays (which are not reduced in this scenario) and the $\gamma$ rays resulting from $\alpha$ capture remain the main challenge -- highlighting the importance of in-situ measurements of these backgrounds. 

In the next section, we investigate the potential to reduce these backgrounds at the reconstruction level (Section~\ref{sec:basicQPix}) and with additional tools such as fiducialization (Section~\ref{sec:fidu}), directionality (Section~\ref{sec:Direction}) or light detection (Section~\ref{sec:light}).

\subsection{Background rejection with Q-Pix clustering}\label{sec:basicQPix}

The first step to discriminate signal and background events is to use the clustering tools described in Section~\ref{sec:Q-PixClustering}, since different particle types may have specific topological signatures and energies. Only $\alpha$ interactions present significant topological differences, whereas 
$\beta$, $\gamma$ (via Compton scattering or pair production), and neutron interactions (via capture or inelastic scattering, producing $\gamma$ rays) all result in the production of single electrons that have identical topology to solar neutrino interactions. Neutrons and $\gamma$ rays can leave additional ``blips" (isolated energy depositions) in the event that could be used to veto. Given the very high event rates of all the other backgrounds, the topological identification of multiple blips for  neutron and $\gamma$ rejection grants in detail consideration and is beyond the scope of this paper.  We investigated the impact of clustering thresholds (CT) on background suppression. 

The $\alpha$ particles, which will have a short very small range in LAr, will leave a very different signature compared to that of single electrons. In our study, we found that $\alpha$ particles  get clustered with a maximum of 4 resets, which is equivalent to the number of resets that a $\sim 1$~MeV electron would deposit in our readout. It is also possible to further use this reset threshold to remove areas of the event spectrum shown in Fig. \ref{fig:spectrum} where the signal is completely buried under background, which is around 3 MeV. For the remaining of the paper, we apply a reset threshold cut of 12, which corresponds to an equivalent deposited energy of 3 MeV, removing all events below this energy. 

\subsection{Fiducialization}\label{sec:fidu}

\begin{figure}[htbp]
\centering
\includegraphics[width=0.9\textwidth]{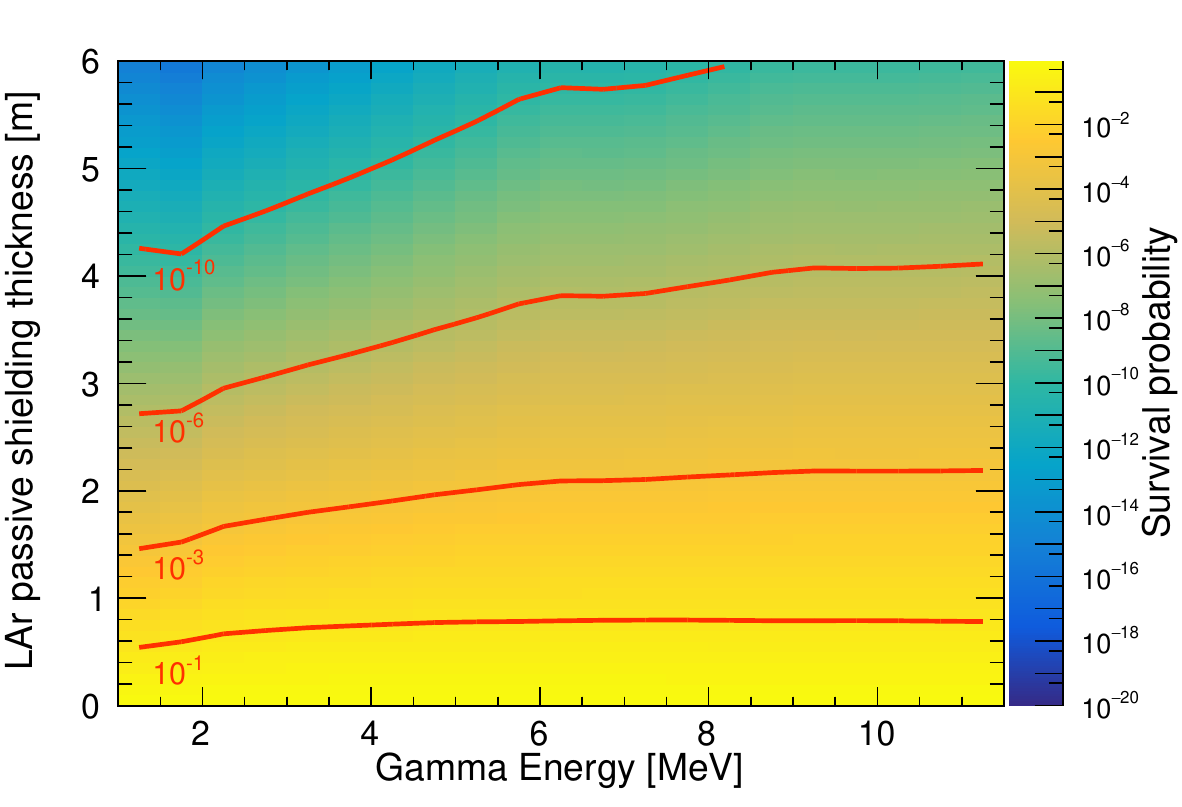}
\caption{\label{fig:fiducialization} 
External $\gamma$ survival probability as a function of initial energy and the thickness of argon used as passive shielding. The $10^{-1}$, $10^{-3}$, $10^{-6}$, $10^{-10}$ contours are shown in red. 
}
\end{figure}

\noindent Since one of the major backgrounds consists of radiation produced outside the detector, we explore the obvious option of fiducializing the argon volume to reduce the impact of cavern $\gamma$ rays and neutron inelastic scattering. Fig.~\ref{fig:fiducialization} shows the external $\gamma$ survival probability as a function of the initial $\gamma$ energy and the distance from the cryostat walls. As expected, $\gamma$ events are suppressed exponentially as a function of the fiducializing distance, while the isotropic nature of the solar neutrino interactions makes the reduction in signal only linear. 

\begin{figure}[h!]
\centering
\includegraphics[width=\textwidth]{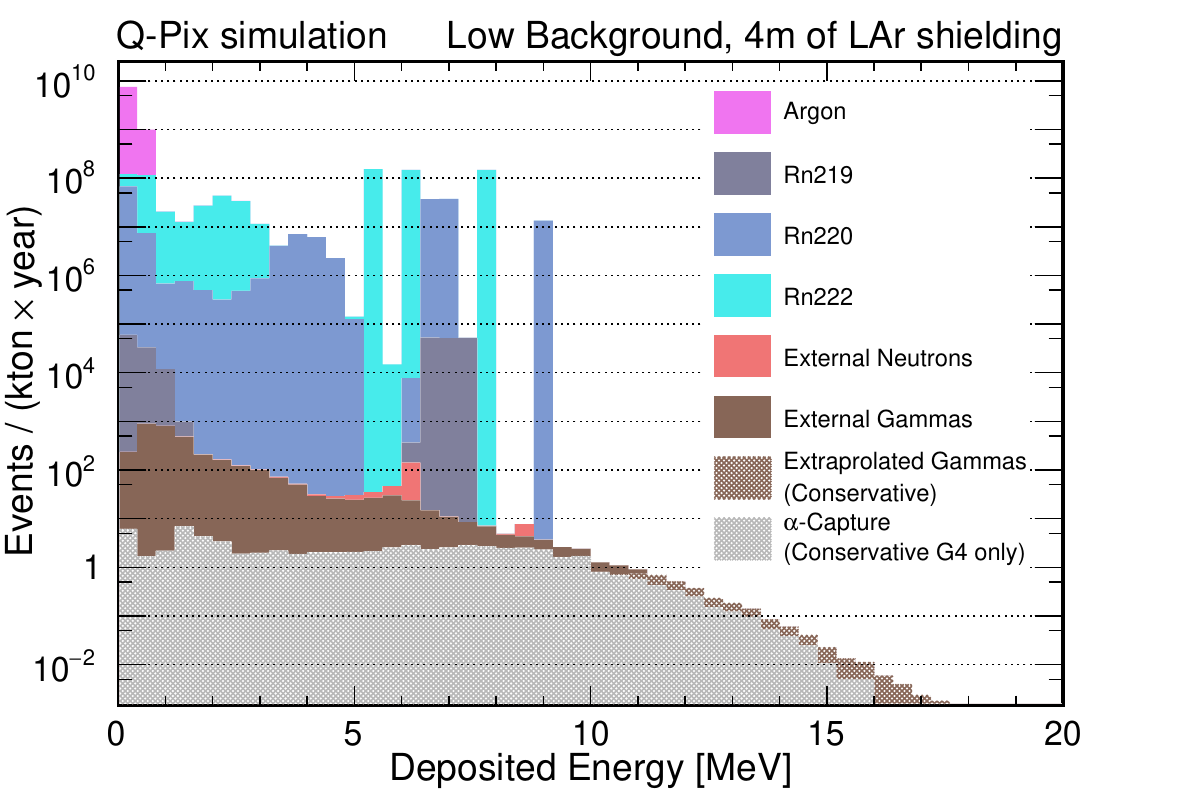}
\caption{\label{fig:spectrum_fid_lowB} 
Deposited energy spectrum of signal and background sources for the low-background scenario using 4 m of LAr as passive shielding.}
\end{figure}

 We provide here an example of the use of the fiducialization tool alone. By considering the straightforward (but aggressive) strategy of using $4$~m of argon as passive shielding, the active volume is reduced to the two innermost TPC volumes (see Fig.~\ref{fig:detector}) of which only $6$~m of height and $54$~m of length remain usable (here we consider $0.5$~m, 
 $1$~m,
 and $1.9$~m of clearances between the detector and the cryostat in the drift, vertical and horizontal axis). This allows the retention of about $20\%$ of the total argon active mass 
 ($\approx 2.3$~kt). The event rejection on the vertical and longitudinal axis can be accomplished precisely by Q-Pix, as the pixels' $4$~mm pitch provides enough spatial resolution for this task. This fiducialization scheme results in the retention of $\sim \mathcal{O}(10^{4})$ solar neutrino events,  $\sim \mathcal{O}(10^{4})$ $\gamma$ rays, $\mathcal{O}(1)$ neutron inelastic scattering events per year.  As can be seen in Fig. \ref{fig:spectrum_fid_lowB}, the stringent fiducialization allows to see much more of the solar neutrino signal over a larger range of energies. 
The hep signal also becomes visible. However, the capability to observe the hep neutrinos will depend on the detailed understanding of the $\gamma$ rays from the $\alpha$-capture on argon and of the external $\gamma$ rays above 11 MeV. It will also require a sufficient energy resolution in order to identify the hep signal in the $\sim$15-18 MeV range. These simple fiducialization strategy can be used as guidance for any  study of solar neutrinos, but specific optimizations will be necessary in any dedicated analyses where multiple background rejection strategies are applied in conjunction, and are thus beyond the scope of this paper.

\subsection{Background rejection with directionality}\label{sec:Direction}
\noindent One key difference between solar neutrino interactions and backgrounds is that neutrinos originate from a known direction: the Sun. While neutrino CC interactions do not retain any directional information from the incoming neutrino, ES events do. A previous study of the detectability of neutrinos from supernovae~\cite{Q-Pix_SN}, showed that Q-Pix enables the reconstruction of the direction of electrons from low-energy ES neutrino interactions. Thus, we explore the possibility of using directionality as a background discrimination tool in the case of solar neutrinos. 

\begin{figure}[htbp]
\centering\includegraphics[width=0.90\textwidth]{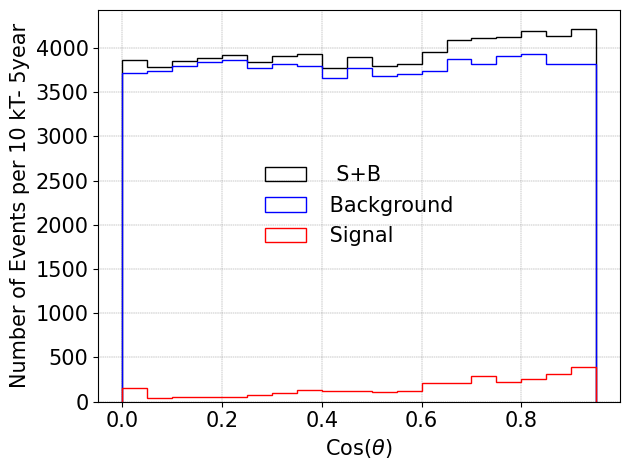}
\caption{\label{fig:lkl} 
Signal (red), background (blue) and total event (black) histograms for the events with reconstructed energy between 6 and 12~MeV in 10kT for a period of 5 years. $\theta$ is the angle between the direction of the sun and the reconstructed primary electron's direction. The low-background scenario and a 4 m fiducialization strategy are assumed here.} 
\end{figure}

In this study, we assume that the Q-Pix clustering threshold is set to 12 resets, enough to suppress all events depositing less than 3 MeV of energy. We reconstruct the directionality of the primary electrons from ES events following the procedure in our previous supernova study with Q-Pix \cite{Q-Pix_SN}.  We define $\theta$ as the angle between the true neutrino direction from the simulation and the reconstructed electron direction. In our study, signal events are generated following the reconstructed angular distribution of the ES reaction, later converted to the $\cos{\theta}$ space. At first approximation, background events follow an isotropic distribution in $\cos{\theta}$. The angular distributions of signal (ES) events and uniform background events are presented in Fig.~\ref{fig:lkl}. We study the signal sensitivity via a likelihood ratio test. The signal and background joint distribution is fitted under the $\mathcal{H}_{0}$ (only background) and the $\mathcal{H}_{1}$ (signal plus background) hypothesis, from which $h_{0}$ and $h_{1}$ likelihoods are respectively obtained. The statistic $-2\ln{(h_{0}/h_{1})}$ of the joint distribution is then compared to that of the background-only distribution and the median sensitivity to the signal is extracted. 

While in the case of supernovae neutrinos directionality was proved to be a powerful tool to point back to the supernova event, we find that its discrimination power in the case of solar neutrinos is severely hampered by the high level of backgrounds. If we combine directionality and a stringent fiducialization (4 m) in the low-background scenario, we find that we can expand sensitivity to solar neutrinos in the 6 to 12  MeV region. Without fiducialization or in our high background scenario, the high level of backgrounds completely disallow the identification of the solar neutrino population on the directionality plot.

\subsection{Background rejection with light detection}
\label{sec:light}

\noindent Light detection in a LArTPC provides additional event information, such as timing, position and energy sensitivity. In this section, we identify areas and tools where a powerful light collection system can confer additional benefits compared to the measurement of charge alone. 

The results are obtained via a Geant4 toy simulation of the full detector geometry where the light production and propagation are simulated in full, but no photosensitive element is explicitly defined. Instead, we assume a photon detection system that provides $100\%$ light coverage at the pixel planes, inline with a proposed innovative solution for a dual charge-light readout using Q-Pix~\cite{bib:ase}, and corresponding to $37\%$ light coverage for the full detector. We assume a conservative $15 \%$ photon detection efficiency, similar to current performances of commercial photosensors~\cite{bib:hpk_vuv}, and a timing resolution better than 10 ns. From these assumptions we estimate the associated background rejection factor. 

We highlight two specific techniques leveraging light detection, namely pulse-shape discrimination between signal and background $\gamma$ rays from $\alpha$-capture events, and delayed light emission from the excited $\ce{^{40}K^{*}}$ produced in CC interactions

\begin{equation}
    \nu_{e} + \ce{^{40}Ar} \rightarrow \ce{^{40}K^{*}} + e^{-} \; .
\label{eq:delayed_gamma}
\end{equation}

\subsubsection{Pulse-shape discrimination}
\label{sec:PSD}
\noindent Fig. \ref{fig:spectrum} highlights the background contributions from the poorly-constrained $\alpha$-capture process reaching up to $17$~MeV in energy and thus interfering with the major portions of $^8$B and hep solar neutrino spectra.

\begin{figure}[ht!]
\centering
\includegraphics[width=0.9\textwidth]{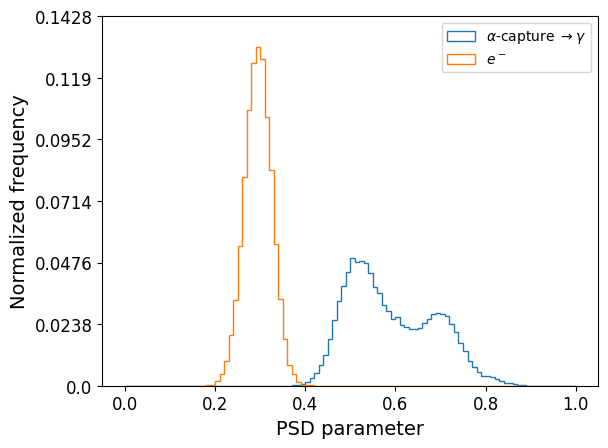}
\caption{\label{fig:alpha_gamma_psd} 
Distribution of the the ratio of fast and slow scintillation light (PSD parameter) for electrons (orange) and $\alpha$-capture $\gamma$ rays (blue).}
\end{figure}

The $\gamma$ rays produced by ionizing $\alpha$ particles capturing on argon present a different scintillation light profile compared to electrons or regular $\gamma$ rays in LAr. When the $\alpha$ particle ionizes the medium before being captured, the initial ionization by the $\alpha$ skews the scintillation distribution for $\alpha$-captures~\cite{KUBOTA1978561}, allowing to use pulse-shape discrimination (PSD) techniques to differentiate $\alpha$-capture $\gamma$ events from solar neutrino events. By comparing the ratio of fast and slow scintillation light, rejection of $\alpha$-capture events is achievable. 
Fig.~\ref{fig:alpha_gamma_psd} shows the distribution of the PSD parameter for both signal (electrons) and background ($\gamma$ rays from $\alpha$ capture) events. This parameter represents the ratio between the fast and slow component of the light, where the integration window for the fast component is set at 50~ns.

By selecting events with a PSD $<$ 0.4, the study shows that 99\% of electron events can be retained while rejecting 99\% of events from $\alpha$-capture events where the $\alpha$ ionizes. While this result is encouraging, our study uncovered limitations in the simulation of the $\alpha$ transport in the range from 1.1 to 8.9 MeV in Geant4. In particular we find that Geant4 reports zero ionization for more than $60\%$ of $\alpha$-capture events. To understand the exact amount of ionization produced by $\alpha$ particles  before they capture as a function of energy, a dedicated measurement of $\alpha$-capture processes in liquid argon will be needed to draw solid conclusions about this harmful background.


\subsubsection{Light coincidence from solar neutrino CC interactions}

\begin{figure}[htbp]
\centering
\includegraphics[width=0.9\textwidth]{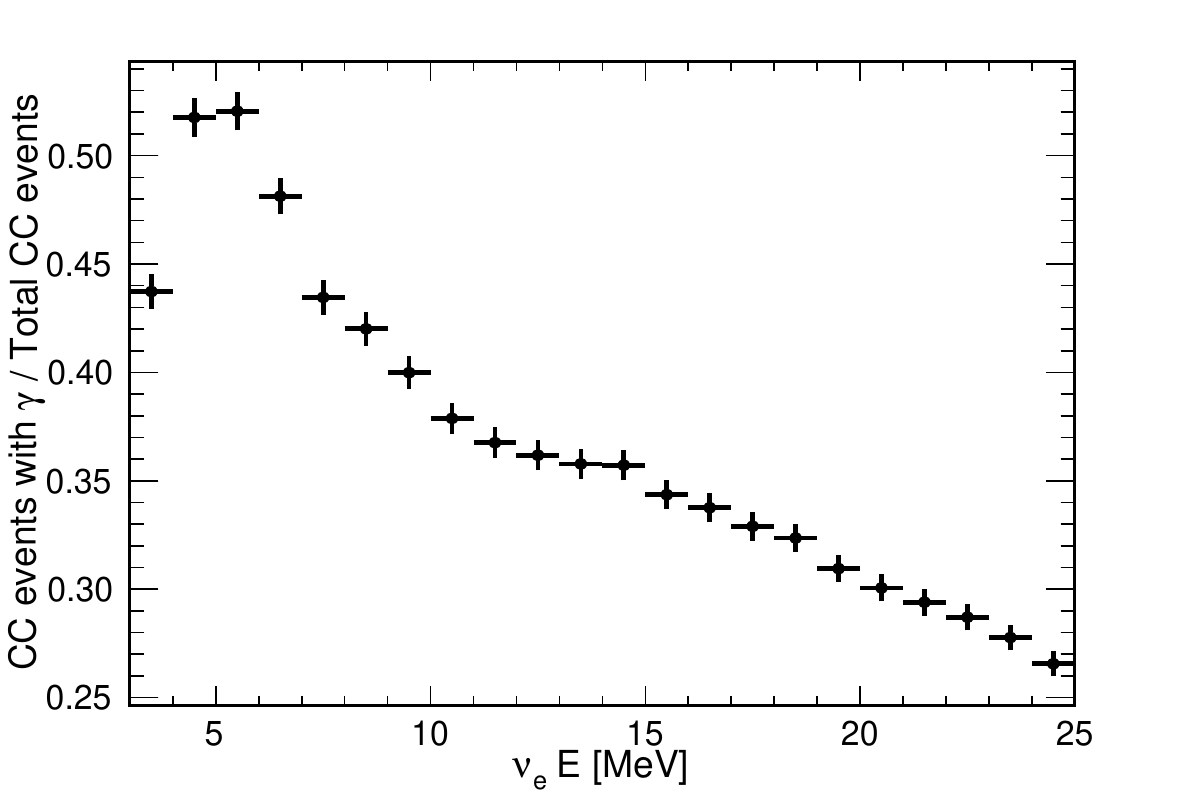}
\caption{\label{fig:delayed_gamma_eff_top} 
Fraction of CC events generating a delayed $\gamma$ as a function of the energy of the incoming solar neutrino.}
\end{figure}

\noindent Charge current interactions from solar neutrinos on argon (\ref{eq:delayed_gamma}) could be isolated from background via the detection of a delayed light flash emitted by the de-excitation of the $\ce{^{40}K^{*}}$ ~\cite{bib:delayed_gamma_1}. Fig.~\ref{fig:delayed_gamma_eff_top} shows the fraction of CC interactions for which the final-state $\ce{^{40}K^{*}}$ atom can de-excite by emitting a 1.64~MeV $\gamma$ with half-life of 336 ns  \cite{bib:delayed_gamma_2}.  
While 336 ns is too short a time to produce distinguishable charge signals, it is long enough to resolve via light detection. 

\begin{figure}[htbp]
\centering
\includegraphics[width=0.99\textwidth]{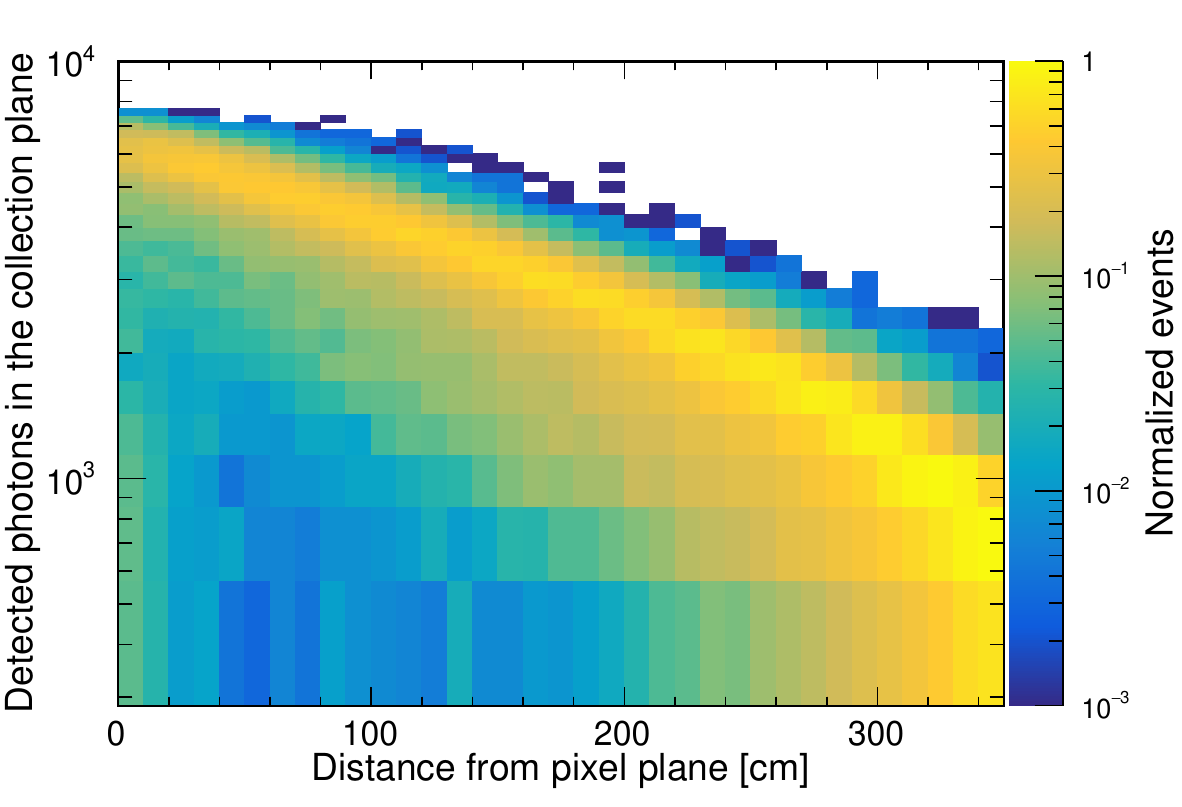}
\caption{\label{fig:delayed_gamma_eff_bottom} 
Photon yield for the delayed $\gamma$ as a function of the distance to the pixel plane.}
\end{figure}

Fig.~\ref{fig:delayed_gamma_eff_bottom} shows the number of detected photons arriving at our assumed light detection system placed on the anode plane for 1.64~MeV $\gamma$ rays isotropically distributed inside the detector. Even for delayed flashes emitted 3.5 meters away from the anode, which is the drift distance in our detector geometry, the number of photons detected is of the order of thousands, building confidence in the feasibility of this technique. We define a time window of 1500 ns to determine a signal coincidence, since such time frame contains more than 99\% of the delayed $\gamma$ flashes. 

\begin{figure}[htbp]
\centering
\includegraphics[width=0.90\textwidth]{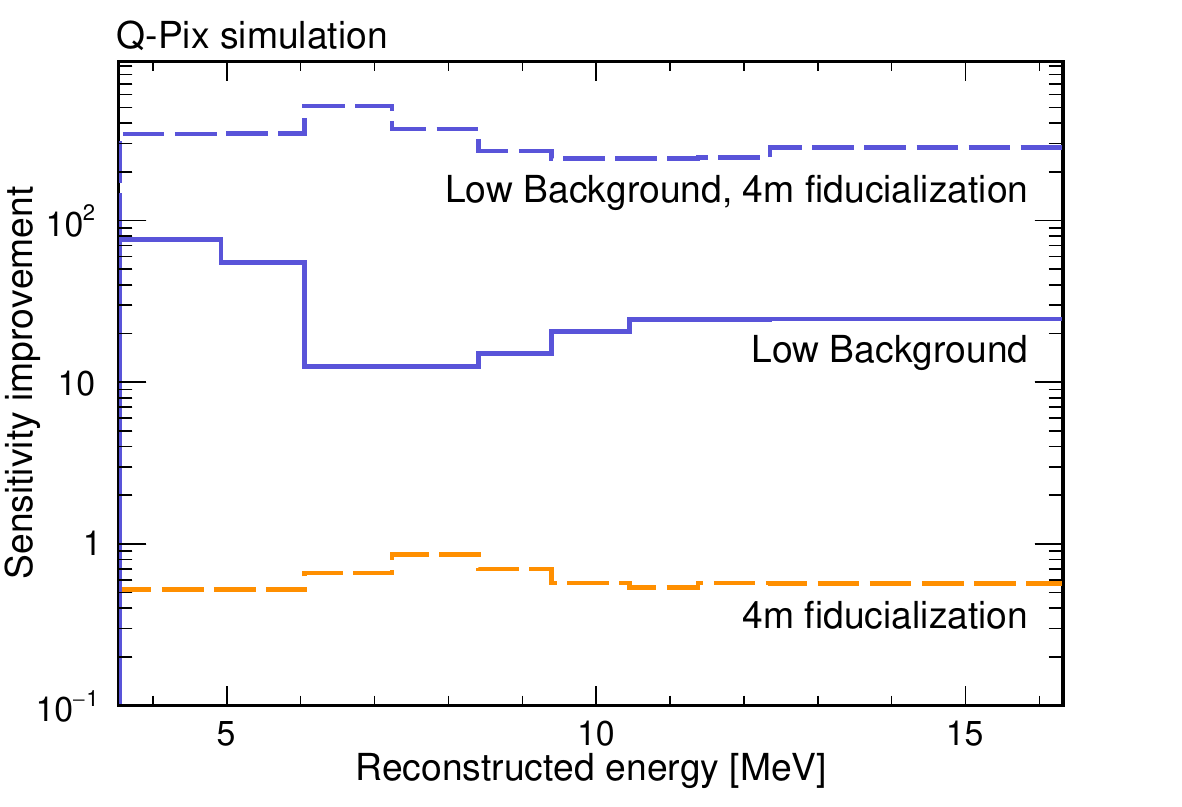}
\caption{\label{fig:cclight_sig} Improvement of background rejection for a solar neutrino measurement in different detector scenarios with respect to the high background case when utilizing the delayed-flash technique and the strong fiducialization. The improvements is calculated as the ratios of the $s/\sqrt{b}$ values.}
\end{figure}

Given the enormous rate of background events, and under the assumption that all background events will produce scintillation light, the viability of this tool depends on the expected number of false positives produced by random background coincidences. For every background source considered in Table~\ref{tab:bgs}, the expected number of events $\mu_{bkg,i}$ in a time window of 1500 ns is computed. $P(\mu_{bkg,i},0)=e^{-\mu_{bkg,i}}$ represents the Poissonian probability of zero events happening in that time window. The probability of zero random background coincidences after an initial event is computed as the product of the individual probabilities, $P(0)=\prod_{i}P(\mu_{bkg,i},0)$. Here, we consider all different sources of signal and background to be independent, and we neglect contributions between different drift volumes. Under our high background scenario and considering the full 10 kton active mass, the probability of a random coincidence is approximately $99\%$. However, in the low-background scenario, the probability of a random coincidence is only 33\%. Further considering $4$~m of liquid argon used as passive shielding, the random coincidence probability drops to 87\% and 0.1\%, respectively for the high background and low-background scenarios. Following the same strategy that was presented in the directionality studies, the Q-Pix clustering algorithm can be used as a pre-selection tool. We set a clustering threshold of 12 resets to select events with energy higher than 3 MeV, allowing the light detection system to be used afterwards to identify a potential delayed $\gamma$ pulse. The number of signal events is computed as the number of CC events passing the CT cut, multiplied by the fraction of events generating a delayed-$\gamma$. The background coincidences are computed as events passing the CT cut (namely, background electrons with energy above $\sim$3 MeV) followed by a light flash produced by any other event. 
Figure \ref{fig:cclight_sig} shows the sensitivity improvement to solar neutrinos (hep and $^8$B CC channels) when utilizing the fiducialization and the coincidence technique for the different detector scenarios. One can see that that applying the fiducialization strategy does not work in the high background scenario, as the background coincidences mimicking the delayed flash are dominated by internal backgrounds. However, when both of these tools are applied in the case of the low-background, fiducialized scenario, sensitivity to solar neutrinos is boosted by more than a factor 100 across a broad range of reconstructed energies despite the rather low total number of solar neutrino events per year.



\subsection{Expected data rates}\label{sec:datarates}

\begin{figure}[htbp]
\centering
\includegraphics[width=0.9\textwidth]{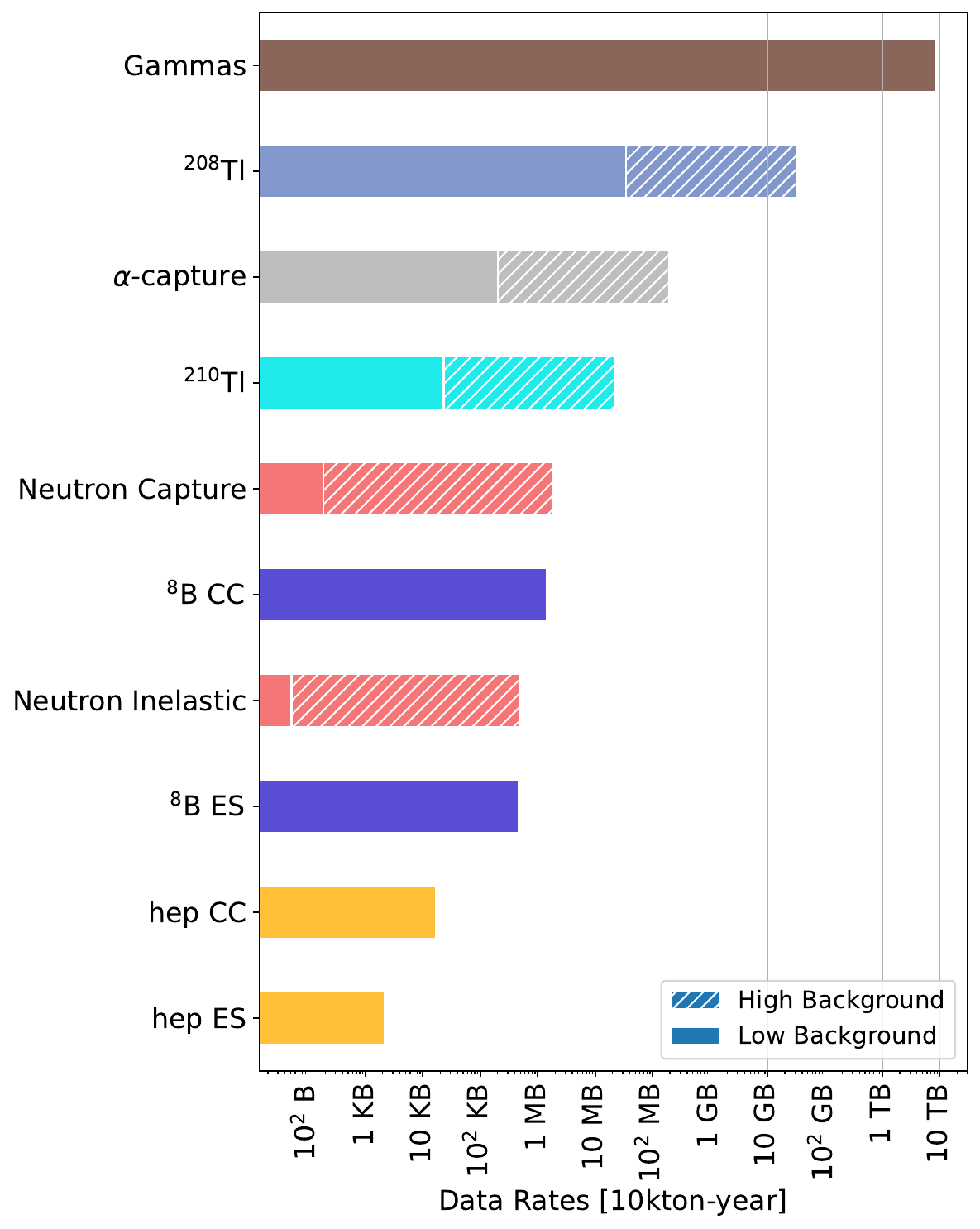}
\caption{\label{fig:qpix_datarates} 
Expected data rates with Q-Pix readout per kt$\cdot$yr for solar neutrino signal events and background processes in the high background scenario (full bar length) and low-background scenario (removing the hatched region) for a CT of 12 resets.}
\end{figure}

\noindent Figure~\ref{fig:qpix_datarates} shows the expected Q-Pix data rates estimated for the solar neutrino events and the leading background sources for the high background and the low-background scenarios after a preselection of a minimum of 12 resets ($\sim 3$~MeV) per event. Data rates in Q-Pix are calculated by counting the number of resets in the events and considering that each reset occupies 8 bytes of memory. The total data rate to store all energy depositions above 3 MeV, needed for an offline solar neutrino analysis, is estimated to be $\sim 1$ TB per year. If recording data below 3 MeV is desired, Q-Pix can store all events with a minimum of 1 reset (i.e., events with deposited energy $\gtrsim 0.15$~MeV), leading to a total data rate of $\sim 1$~PB per year, still a manageable amount.

In comparison, the expected data size for an event readout from a wire or CRP-based LArTPC is much larger.  Reading out a full wire-based 10-kton DUNE module occupies about $6.5$~GB if waveforms are recorded for $5.4$~ms~\cite{DUNE:2020lwj} and $8$~GB if data from a CRP readout is recorded for $4.25$~ms~\cite{bib:vd_tdr}.  
We estimate that the rate of events worth recording for a solar analysis -- signals and backgrounds that deposit 3 MeV or more -- is around 300 Hz. Such a high rate implies that at least one potential solar event is present in each drift time of a wire- or CRP-based detector. Continuously reading out such detectors for a year would generate of the order of $10^{5}$~PB of data. If advanced trigger techniques can allow to record information of a single APA or CRP instead of the full detector, the data rate could be reduced by two orders of magnitude -- still significantly higher than the expected rates generated by the Q-Pix continuous readout.

\section{Conclusions} 
\noindent We have studied the feasibility of solar neutrino detection using a LArTPC equipped with Q-Pix technology. Our study outlines the main challenges in detecting solar neutrinos that must be addressed by future large-scale noble element detectors, such as DUNE~\cite{DUNE_FD_rq}, SOLAIRE~\cite{price_solaire_2024}, and -- with the due caveats pertaining to the change of target nucleus -- XLZD~\cite{lxzd}. 

Our analysis includes an extensive evaluation of potential background sources that could affect solar neutrino detection in an underground LArTPC detector. Internal radioactive backgrounds originating from bulk argon and radon decays make it difficult to detect solar neutrinos below a deposited energy of $5$~MeV, as background rates exceed signal rates by nearly ten orders of magnitude.

Beta decays and $\gamma$ rays from radioactive processes — including rare decay — create an irreducible background that exceeds the solar neutrino signal by five to eleven orders of magnitude, depending on the energy range. This conclusion remains unchanged even when accounting for unmodeled mitigation effects such as ion drift or including additional sources of radioactivity from detector components like readout support structures. Although a low-background scenario significantly reduces internal background rates, they still remain too high to enable the study of solar neutrinos below $5$~MeV.

Above $5$~MeV, the dominant background sources are photons originating externally to the detector—which can only be mitigated through fiducial volume cuts—and photons generated by $\alpha$-capture processes in argon. These $\alpha$ particles arise from radon decay chains and could be reduced in a low-background detector scenario, with further suppression possible using techniques like pulse shape discrimination.

However, the two primary background processes above $5$~MeV are poorly constrained due to the limited availability of measurements of $\gamma$ fluxes in the detector cavern and $\alpha$-capture cross sections on argon. We explored the possibility of isolating hep neutrinos and obtaining a high-statistics sample of $^8$B neutrinos in a low-background scenario, provided that strict fiducialization and flash-coincidence techniques are employed to identify solar events. Nonetheless, any definitive conclusion regarding hep neutrino discovery or detailed $^8$B neutrino studies depends critically on precise measurements of cavern $\gamma$-ray emission and the $\alpha$-capture process.

If future measurements confirm that $\gamma$ rates and $\alpha$-capture processes fall within the assumed order of magnitude, we have shown that offline analysis tools can significantly enhance the potential for solar neutrino studies in a LArTPC equipped with a pixelated readout and an effective light detection system. 

Finally, we emphasize the necessity of a continuous readout system for solar neutrino detection. As demonstrated throughout this study, the background rates — despite caveats in some assumptions — are so high that traditional triggering approaches in LArTPCs are unlikely to be viable. Q-Pix technology, which enables significantly reduced data rates, emerges as a strong candidate to support continuous readout capabilities essential for these studies. 


\begin{acknowledgments}
This material is based upon work supported by the
U.S. Department of Energy, Office of Science, Office of
High Energy Physics Award No. DE-0000253485 and
No. DE-SC0020065 and by the STFC grant ST/W003945/1. S. Kubota is supported by the Ezoe Memorial Recruit Scholarship, Mitsui Group 350th Anniversary Project, ONE CAREER Inc. and PKSHA Technology Inc.  G. Ruiz and S. S\"oldner-Rembold have received funding from the European Union’s Horizon 2020 Research and Innovation programme under GA no 101004761. J.B.R.Battat is supported by the Gordon and Betty Moore Foundation through Grant GBMF11565 and grant DOI https://doi.org/10.37807/GBMF11565. O. Seidel is supported by the High Energy Physics Integrated Circuits Fellowship under Award No. DE- SC0022296
\end{acknowledgments}

\newpage 
\bibliographystyle{apsrev4-2}
\bibliography{solar-bib}

\end{document}